# COVID-19 response needs to broaden financial inclusion to curb the rise in poverty

Mostak Ahamed♦ and Roxana Gutiérrez-Romero♣

29 MAY 2020

ABSTRACT

The ongoing COVID-19 pandemic risks wiping out years of progress made in reducing global poverty. In this paper, we explore to what extent financial inclusion could help mitigate the increase in poverty using cross-country data across 78 low- and lower-middle-income countries. Unlike other recent cross-country studies, we show that financial inclusion is a key driver of poverty reduction in these countries. This effect is not direct, but indirect, by mitigating the detrimental effect that inequality has on poverty. Our findings are consistent across all the different measures of poverty used. Our forecasts suggest that the world's population living on less than $1.90 per day could increase from 8% to 14% by 2021, pushing nearly 400 million people into poverty. However, urgent improvements in financial inclusion could substantially reduce the impact on poverty.

♦ University of Sussex Business School, University of Sussex, Brighton, UK. M.Ahamed@sussex.ac.uk.

♣ Queen Mary University of London, School of Business and Management, London, UK. r.gutierrez@qmul.ac.uk.

Centre for Globalisation Research (GCR) working paper series, School of Business and Management, QMUL, London, UK.

1. INTRODUCTION

The ongoing COVID-19 pandemic will have a profound health and economic impact, particularly in the developing world. Millions of people in these economies are employed in the informal sector often without regular access to welfare or pension rights. With the social distancing and lockdown measures implemented to control the spread of COVID-19 millions of people suddenly lost their livelihoods and can no longer rely on their daily earnings to survive. Urgent cash transfers and labour initiatives have been implemented in 181 countries to try mitigating some of the immediate economic impacts of the pandemic (Gentilini et al., 2020). However, other short-term and medium-term policies will be needed to help households receive government transfers and build financial buffers to spread resources over the likely prolonged crisis. Globally, there are 1.7 billion adults without an account at a financial institution or a mobile money provider (World Bank, 2018). In this context, it is more important than ever to understand to what extent financial inclusion could contribute to reducing poverty, and how.

In this paper, we address both questions. We estimate to what extent changes in poverty have been the product of improvements in income or distributional gains. To this end, we follow the poverty decomposition approach proposed by Datt and Ravallion (1992) using cross-country data on poverty and inequality for 78 low- and lower-middle-income countries over the last two decades. We take these series from PovcalNet, a dataset provided by the World Bank (Atamanov et al., 2018). Our contribution is to estimate to what extent financial inclusion has played a direct effect on poverty reduction. We also consider whether financial inclusion has had an indirect effect by mediating the impact that inequality and growth have on poverty. To this end, we construct a multidimensional index of financial inclusion proposed by Ahamed and Mallick (2019) using the Financial Access Survey over the period between 2004 and 2018 (International Monetary Fund, 2019). In addition to this overall index of financial inclusion, we also provide an index of financial outreach and usage. In the Appendix, we present such indices by country during the 2004-2018 period.

Our paper offers four key findings. First, we show that over the last two decades, poverty has been relatively unresponsive to economic growth and has been deeply affected by the still high levels of inequality. These findings are in line with earlier literature (Datt & Ravallion, 1992; van der Weide & Branko, 2018). Second, we contribute to the literature by showing that financial inclusion does not directly reduce poverty; however, it strongly reduces the detrimental effect that inequality has on poverty. In other words, unlike other recent cross-country studies, we show that financial inclusion is a key element in reducing poverty, even in low- and lower-middle-income countries (e.g. Goksu et al., 2017; Park & Mercado, 2018). Third, we also show that increasing both the outreach and usage dampens the detrimental effect that inequality has on poverty. Our results are robust to using the three components of poverty measures: incidence, intensity, and inequality among the poor. Our findings are also robust to using alternative measures of financial inclusion taken from the World Bank's Global Findex. Fourth, our forecast analysis suggests that the world's population living on less than $1.90 per day could increase from 8% to 14% by 2021, pushing 400 million people into poverty if no urgent and adequate measures are implemented. However, our forecast analysis also suggests that, with improvements in financial inclusion, poverty increases could be curbed.

The remainder of the paper is structured as follows. The next section provides a brief overview of the literature. Section 3 describes the data and discusses our estimation strategy. Section 4 presents the empirical results, while Section 5 shows the forecast analysis. Section 6 presents our conclusion.

2. LITERATURE REVIEW

The empirical literature on financial inclusion falls broadly into three main categories. The first one uses randomised control experiments to ascertain the impacts of offering financial services or improving outreach to individuals, households, and firms. The second strand uses quasi-experiments and case studies using mostly *ad-hoc* measures of financial inclusion. The third strand analysis evaluates the impacts of financial inclusion using cross-country aggregate analysis. What we can learn from this extensive literature is that the role of financial inclusion in poverty reduction is far from conclusive.

During the 1980s and the 1990s, it was widely believed that financial inclusion, particularly in the form of providing micro-credits, could be vastly beneficial for poverty reduction (Morduch, 1999; Yunus, 2013). However, recent evidence of the micro-credit revolution stemming from randomised control trials is more nuanced. Micro-credits do improve people ability to earn a living and help some to create and expand small businesses, but the evidence on poverty reduction is negligible. Systematic reviews of micro-credit have failed to find positive effects on household income (Duvendack, Palmer-Jones, & Vaessen, 2014; Stewart et al., 2012), including a meta-analysis from Grameen Bank micro-credits (Yang & Stanley, 2014). Although there are no dramatic changes in poverty reduction, there is no evidence either that these micro-credits lure vulnerable people into indebtedness, as some isolated anecdotal evidence might suggest (Banerjee, Duflo, Glennerster, & Kinnan, 2015).

The literature has also focused on understanding the extent to which a broader financial inclusion can be pro-poor and what sort of broader inclusion is needed. Is it merely expanding the outreach, so that poor people can access financial services, or is it more on also focusing on increasing financial usage? The evidence is again, inconclusive. From case studies, we know that increasing financial outreach can be beneficial for poverty reduction, even when outreach expansion might be motivated by political reasons (Cole, 2009). However, questions remain as to whether this is the most effective way of resource allocation instead of, for instance, direct cash transfers. For example, the largest mandated bank branch expansion established in rural areas in India during 1969-1990 helped to reduce poverty (Burgess & Pande, 2005). Still, the bank loan default rate was 40%, and questions remain about its cost-effectiveness, relative to potential alternative programmes.

Over the last decade, financial inclusion has also focused not only on improving access to credits, but also on broadening access to financial services, such as savings, insurance, and mobile banking (Cai, Chen, Fang, & Zhou, 2009; Dercon, 2005; Flory, 2018). The evidence from randomised control trials of improving household income, thanks to providing access to micro-savings and insurance, is promising. There is also evidence that expanding access to savings can particularly benefit those users that have been typically constrained and reduce sharp gender inequalities. For instance, the first randomised control trial of this kind provided access to non-interest-bearing bank accounts to young women and men in Kenya (Dupas & Robinson, 2013). The experiment showed that, despite hefty withdrawal fees, the majority of women used the accounts, and they were able to save more and to increase their investment and expenditures than men. This study suggests that women, particularly in rural areas, face negative private returns on money if they cannot find secure forms of saving. Similar findings have been found in Nepal and Malawi (Flory, 2018; Prina, 2015).

These recent randomised control trials suggest that financial inclusion might not only help poor people have more productive investments, smooth their consumption from idiosyncratic or local shocks but also help to reduce inequality. Recent quasi-experimental and case studies in developing countries such as in China, India, Nigeria and Ghana also suggest that increasing financial inclusion, in the form of increasing outreach and usage, can help to reduce household vulnerability to poverty, particularly in those with financial services in distant places (Churchill & Marisetty, 2020; Dimova & Adebowale, 2018; Koomson, Villano, & Hadley, 2020; Li, 2018). But a remaining question is whether financial inclusion helps reducing poverty directly or via its impact on reducing income and gender inequalities. From cross-country studies, there is some mixed evidence. For instance, Goksu, Deléchat, Newiak, and Yang (2017), using a micro-data set across 140 countries, found a non-linear relationship between financial inclusion and inequality. Their findings suggest that in lower stages of development, only a small group, the wealthy, benefit from financial inclusion progress, but with a broader level of financial inclusion gradually all other groups benefit. Similarly, using cross-country analysis, Park and Mercado (2018) show that financial inclusion is positively associated with lower levels of poverty in high- and upper-middle-income economies, but not in middle-low and low-income economies. This mixed evidence is perhaps unsurprising, given that high-income economies have a broader welfare system and better regulatory conditions that can further the impact of financial inclusion. However, this earlier analysis needs to be broadened to understand the main factors that drive poverty changes and how financial inclusion might affect poverty, whether directly or indirectly. In this respect, the literature has found two challenges. The first regards how the multidimensional aspects of financial inclusion should be measured, and the second, how to estimate whether financial inclusion affects poverty directly or indirectly by dampening the detrimental effects of inequality.

Over the last decade, the literature has proposed several different measures of financial inclusion, mainly drawing from individual financial surveys, or drawing from the global financial surveys conducted by the World Bank or the IMF (Ahamed & Mallick, 2019). The World Bank has recently made available the Global Financial Inclusion database, which provides information on more than 850 indicators across 151 economies, focusing on the demand side of financial services (Demirgüç-Kunt, Klapper, Singer, Ansar, & Hess, 2018). A main constraint of this database is its periodicity, only available for 2011, 2014, and 2017, which does not allow for an extended comparative analysis across countries over time. An alternative source is the Financial Access Survey, 2004-2018 gathered by the IMF (International Monetary Fund, 2019). This annual series offers the largest global supply-side data on financial inclusion, including data on access to and usage of financial services by both firms and households that are comparable across countries and over time.

As reviewed here, all strands of the empirical literature are equally relevant for policy analysis. From the experimental literature and quasi-experimental case studies, we learn that financial inclusion needs to consider more than micro-credits. We also learn that financial inclusion might not reduce poverty directly, but indirectly by reducing inequalities in financial access and by broadening financial usage among typically disadvantaged groups. Cross-country analysis has a different advantage. They allow using the same measure of financial inclusion to make comparative analysis across countries and over time. Cross-country analysis can also estimate the likely direct and indirect impacts of financial inclusion on poverty.

Next, we take advantage of the poverty-growth-inequality decomposition method proposed by Datt and Ravallion (1992) to understand whether financial inclusion affects poverty reduction directly or indirectly. At the aggregate level, poverty from one period to the next might change as a result of changes in the Gross Domestic Product (GDP), ceteris paribus, or whether GDP is distributed any differently. These simple poverty decompositions, theoretically underpinned by Lorenz Curve principles, can be empirically estimated. Extensive research has shown that inequality is in particular detrimental for poverty reduction since increases in GDP are often captured by middle or upper classes, with a limited trickledown effect for the poor (Gutiérrez-Romero & Méndez-Errico, 2017; Ravallion, 2005; van der Weide & Branko, 2018). In our empirical analysis, we extend these decomposition regressions to include the potential role of financial inclusion in poverty reduction. Our first hypothesis is whether financial inclusion might affect poverty reduction directly. Our second hypothesis is that financial inclusion might reduce poverty indirectly by mitigating the detrimental effect that inequality has on poverty. To test our two key hypotheses, we construct a financial inclusion index along two associated dimensions of financial outreach and financial usage, as explained next.

## 3. EMPIRICAL STRATEGY: DATA AND METHODOLOGY

We empirically examine the link between income inequality, financial inclusion, and poverty. In doing so, we follow Datt and Ravallion (1992) in decomposing poverty changes as a result of growth effects or distributional changes, as in equation (1).

$$P_f - P_i = \underbrace{G(f,i;r)}_{\text{Growth}} + \underbrace{D(f,i;r)}_{\text{Redistribution}} + \underbrace{R(f,i;r)}_{\text{Residual}} \qquad (1)$$

where G(.); D(.) and R(.) stand for growth, distribution, and residual components. If we keep the mean income constant at a reference level, the growth component of a change in poverty can be defined due to a change in the mean income; likewise, the redistribution component can be defined due to a change in the Lorenz curve. The residual component can be construed as the difference between the evaluated growth (redistribution) components and the poverty change when the mean income or the Lorenz curve remained unchanged over the period; then, the residual would be zero (Freije, 2014).

Using a sample of low- and lower-middle-income countries for the 2004-2018 period, we estimate a general model, as follows:

$$\Delta P_{it} = \alpha_i + \beta \Delta \text{Gini}_{it} + \gamma \text{GDP Growth}_{it} + \varepsilon_{it} \qquad (2)$$

where $\Delta P$ stands for changes in headcount poverty (or poverty gap, or poverty gap squared, or Watts index). $\Delta$Gini refers to the change in inequality (proxied by the Gini coefficient), and GDP growth is the real GDP growth rate. As mentioned in the literature review, apart from inequality and GDP growth, inclusive financial development might also influence the lives of the poor. Therefore, we have augmented eq. (2), by adding the financial inclusion indicators, as follows:

$$\begin{aligned}\Delta P_{it} = &\alpha_i + \beta \Delta \text{Gini}_{it} + \gamma \text{GDP Growth}_{it} + \delta \Delta(\text{Financial inclusion})_{it} \\ &+ \varphi \Delta \text{Gini}_{it} \times \Delta(\text{Financial inclusion})_{it} + \varepsilon_{it}\end{aligned} \qquad (3)$$

where Financial inclusion is a multidimensional index of inclusive financial development for country $i$ at time $t$. To examine the role of Financial inclusion on the relationship between inequality and poverty, we introduce the interaction term $\Delta$Gini×Financial inclusion.[1] A negative φ coefficient would imply that a country with a higher inequality and a greater inclusive financial development reduces poverty.

---

[1] As discussed in Section 4, we also include an interaction between the index of financial inclusion and growth, but the interaction is statistically insignificant.

*3.1. Data sources*

We compile data from several sources. Data on poverty measures (using the $1.90 a day poverty line), such as headcount ratio, poverty gap, poverty gap squared, Watts index, and income inequality (the Gini coefficient) come from the March 2020 World Bank survey round of the PovcalNet database. PovcalNet is the main repository for empirical poverty research.[2] It contains consumption/income data from over 2 million randomly sampled households in about 164 countries from 1981 to 2018. Since this database uses Purchasing Power Parity (PPP) conversion metrics from the 2011 International Comparison Program, it makes it possible to undertake a cross-country analysis.

Data on the real GDP growth rate for the 2004-2019 period with the forecasted series for the year 2020 and 2021 are from the International Monetary Fund (IMF). To construct the financial inclusion index, we collect information from the IMF's Financial Access Survey (FAS) database. As an alternative indicator of financial inclusion, we collect information on account ownership at a financial institution from the World Bank's Global Findex database.

*3.2 Constructing the financial inclusion index*

The recent literature has proposed to measure progress made in financial inclusion by focusing on two dimensions: financial outreach and financial usage (Ahamed & Mallick, 2019; Amidžic, Massara, & Mialou, 2014). Following this literature, in this paper, we construct an overall index of financial inclusion index, denoted by FII, and also two sub-indices on financial outreach and financial usage.

We construct an index of financial usage to account for the depth of the financial access. This index includes the number of bank accounts per 1,000 people.[3] We also construct an index of financial outreach, which is intended to capture physical proximity to the physical point of financial services, as this dimension is considered to be one of the most important impediments to inclusive financial development (Allen et al. 2014). Therefore, our index of financial outreach includes the pervasiveness of the outreach of the financial sector in terms of bank branches and ATMs. To construct this outreach index, we create four sub-indices using the demographic and geographic penetration of bank branches and ATMs. We use the number of bank branches (Fo1) and the number of ATMs (Fo2) per 100,000 people to capture demographic penetration of bank branches. We also use the number of bank branches (Fo3) and the number of ATMs (Fo4) per 1,000 square kilometres to capture the geographic penetration of bank branches.[4]

---

[2] http://iresearch.worldbank.org/PovcalNet

[3] As the data on the number of people having bank accounts is limited, we use the number of accounts per capita. In the latter case, double counting cannot be eliminated if a person has multiple accounts (for more, see Beck, Demirguc-Kunt, & Martinez Peria, 2007). Note that data on all indicators in constructing an financial inclusion index is based on commercial banks.

Lastly, we construct an index of overall financial inclusion. To capture the multidimensional aspects of financial inclusion, we use principal component analysis (PCA) which helps us to weight different dimensions of this index, as follows:[5]

$$\text{FII} = \sum_{i=1}^{n} w_{ij} X_i \qquad (4)$$

where $w_{ij}$ are the component's loadings or weights, and $X_i$ are the original variables. First, we apply PCA to estimate the financial outreach dimension from a group of four sub-indices (Fo1, Fo2, Fo3, and Fo4), as discussed above.[6] Second, we apply PCA again to estimate the financial inclusion index by using the financial outreach and usage as causal variables.[7] In PCA, the first principal component is the single linear combination of all indicators that explains most of the variation. In constructing the FII, we find that the first PC explains about 89% of the corresponding sample variance with an eigenvalue of more than one, that is, 1.79.

We are interested in financial inclusion and its associated dimensions to examine their role in the relationship between inequality and poverty. Therefore, to ease the interpretation, we normalise FII and assign each country in a 0-1 scale: 1 refers to greater inclusive financial development.

---

[4] This approach follows Beck, Demirguc-Kunt and Martinez Peria (2007).

[5] See Tetlock (2007) for more on principal component analysis.

[6] Using PCA, we created a financial outreach dimension before running it on the financial outreach and usage dimensions. The first PC explains about 73% of the variations with the eigenvalue of 2.91.

[7] We winsorise each indicator at the 95th percentile levels to reduce the influence at the upper tail. We also normalise each indicator to have values between zero and one to ensure the scale in which they are measured is immaterial.

Figure 1 shows the overall FII around the globe. In the Appendix, in Tables A.1, A.2, and A.3, we provide the FII index for each country for which there is available data, including for the overall financial inclusion index, as well as the associated dimensions of outreach and usage from 2004 to 2018. In Table A.1, we have also included a category of whether the country belongs to a low, lower-middle, upper-middle, or high-income. The distinction is important because we exclude those upper-middle and high-income countries from our econometric analysis.

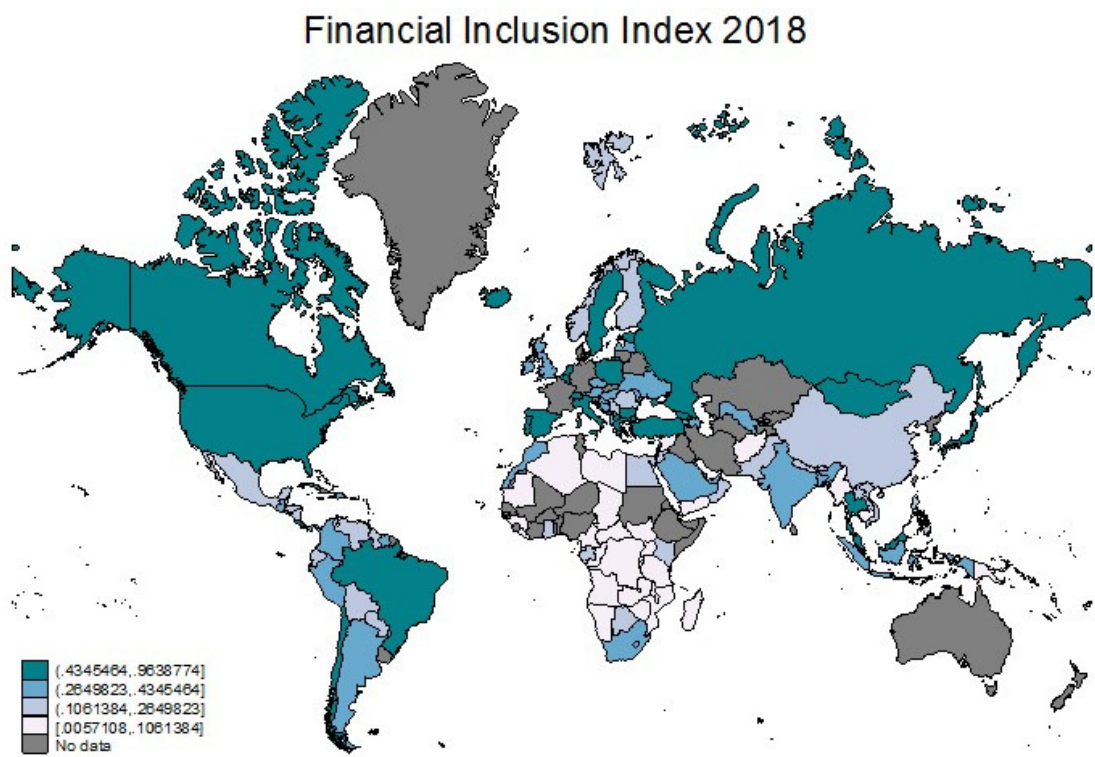

Figure 1: Financial inclusion around the globe.
Source: Own estimates using 2019 Financial Access Survey (IMF, 2019).

*3.3 Descriptive statistics*

After merging PovcalNet data with financial inclusion and GDP growth rate, we end up with 933 observations for a sample of 78 low- and lower-middle-income economies for the 2004-2018 period. Table 1 reports the descriptive statistics of all variables: Panel A shows the summary statistics of the variables in levels, whereas changes are reported in Panel B. The average headcount is 27% with a standard deviation of 25%. A higher headcount means that over one-fourth of the population in these countries lives with under $1.9 a day. The average Gini index of 41% indicates a sizable inequality in the sample economies. The average real GDP growth rate of 4.7% with a standard deviation of 3.6% indicates a heterogeneous economic growth. Likewise, the financial inclusion index of 0.13, with a standard deviation of 0.12, also implies a substantial variation in the extent of inclusive financial development in these countries. Higher standard deviations in financial outreach and usage dimensions also suggest a significant variation in terms of banking sector outreach and usage of financial services in these low- and lower-middle-income countries. In terms of account ownership, this suggests that, overall, 33% of people have an account at a financial institution where men have more (7%) access to financial services, when compared to women.

Table 1. *Summary statistics*

|  | Mean | Median | Standard deviation | Min | Max | NxT |
|---|---|---|---|---|---|---|
| Panel A | Summary statistics in levels | | | | | |
| Headcount | 0.271 | 0.173 | 0.25 | 0.000 | 0.941 | 933 |
| Poverty gap | 0.106 | 0.05 | 0.125 | 0.000 | 0.636 | 933 |
| Poverty gap squared | 0.057 | 0.018 | 0.078 | 0.000 | 0.469 | 933 |
| Watts index | 0.165 | 0.065 | 0.213 | 0.000 | 1.246 | 933 |
| Gini | 0.412 | 0.405 | 0.078 | 0.240 | 0.633 | 933 |
| GDP growth rate | 0.047 | 0.048 | 0.036 | -0.077 | 0.139 | 933 |
| Financial inclusion index | 0.129 | 0.082 | 0.122 | 0.000 | 0.592 | 933 |
| Financial outreach | 0.107 | 0.064 | 0.106 | 0.000 | 0.521 | 933 |
| Usage | 0.156 | 0.111 | 0.163 | 0.000 | 1 | 933 |
| Account ownership | 0.329 | 0.313 | 0.18 | 0.033 | 0.93 | 391 |
| Account ownership (Male) | 0.366 | 0.361 | 0.183 | 0.032 | 0.908 | 391 |
| Account ownership (Female) | 0.293 | 0.277 | 0.185 | 0.011 | 0.95 | 391 |
| Panel B | Summary statistics in changes | | | | | |
| ΔHeadcount | -0.007 | 0.000 | 0.044 | -0.388 | 0.292 | 856 |
| ΔPoverty gap | -0.003 | 0.000 | 0.025 | -0.259 | 0.206 | 856 |
| ΔPoverty gap squared | -0.002 | 0.000 | 0.018 | -0.234 | 0.156 | 856 |
| ΔWatts index | -0.006 | 0.000 | 0.048 | -0.61 | 0.425 | 856 |
| ΔGini | -0.001 | 0.000 | 0.022 | -0.209 | 0.255 | 856 |
| ΔFinancial inclusion index | 0.005 | 0.004 | 0.021 | -0.238 | 0.108 | 856 |
| ΔFinancial outreach | 0.005 | 0.003 | 0.02 | -0.15 | 0.137 | 856 |
| ΔUsage | 0.005 | 0.004 | 0.037 | -0.496 | 0.152 | 856 |
| ΔAccount ownership | 0.02 | 0.000 | 0.057 | -0.073 | 0.323 | 391 |
| ΔAccount ownership (Male) | 0.022 | 0.000 | 0.061 | -0.066 | 0.346 | 391 |
| ΔAccount ownership (Female) | 0.019 | 0.000 | 0.056 | -0.081 | 0.335 | 391 |

Figure 2 illustrates the relationship between poverty, financial inclusion, and income inequality. This figure shows that countries with more inclusive financial development are associated with lower levels of poverty (Panel A) and lower levels of inequality (Panel B).

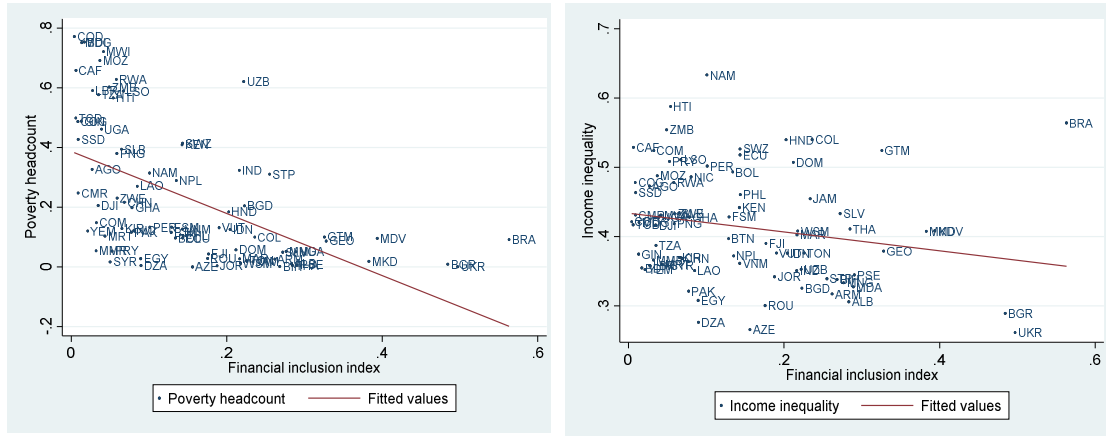

                    Panel A                                     Panel B

Figure 2: Financial inclusion, poverty, and inequality.

## 4. RESULTS

In this section, we first discuss the estimation results of eq. (2) and then move on to discuss the role of financial inclusion in the relationship between inequality and poverty, as in eq. (3). We use panel fixed effect specifications, with robust standard errors clustered at the country level.

      We analyse the observed poverty changes by applying decomposition techniques that separate and gauge the influence of changes in economic growth and inequality. Note that the decomposition technique does not prove causality, but does account for the size of different components and their contributions to changes in poverty (Freije, 2014). As mentioned before, we use Datt-Ravallion decomposition, which measures the extent of poverty change attributable to changes in income growth or distribution of income. Table 2 reports the results of the decomposition. We use four different variants of poverty measure as the dependent variable: ∆Headcount, ∆Poverty gap, ∆Poverty gap squared, and ∆Watts.

      The results show that redistribution (change in Gini) is a dominant effect in increasing poverty. That is, inequality is associated with higher levels of poverty in low- and lower-middle-income economics. On the other hand, economic growth is not a dominant factor in explaining the changes in either of the poverty measures. Apart from column 1, economic growth seems to reduce poverty, but the coefficient is statistically insignificant. Overall, the results imply that poverty changes are mostly driven by changes in inequality or redistribution, but economic growth might be able to hinder increases in poverty.

Table 2. *Growth-redistribution decomposition of poverty*

| Dependent variable | ΔHeadcount | ΔPoverty gap | ΔPoverty gap squared | ΔWatts |
|---|---|---|---|---|
| | 1 | 2 | 3 | 4 |
| ΔGini | 0.618* | 0.501*** | 0.383*** | 1.044*** |
| | [0.326] | [0.178] | [0.124] | [0.348] |
| GDP growth rate | 0.011 | -0.004 | -0.005 | -0.011 |
| | [0.037] | [0.022] | [0.015] | [0.041] |
| Constant | -0.007*** | -0.003** | -0.001* | -0.004* |
| | [0.002] | [0.001] | [0.001] | [0.002] |
| Observations | 856 | 856 | 856 | 856 |
| Adjusted $R^2$ | 0.0977 | 0.199 | 0.227 | 0.23 |
| Number of country | 77 | 77 | 77 | 77 |
| Country fixed effects | Yes | Yes | Yes | Yes |
| Robust standard error cluster | Country | Country | Country | Country |

The dependent variables are changes in *headcount, poverty gap, poverty gap squared, and Watts* index. *ΔGini* is the change in income inequality.

***, **, and * indicate statistical significance at the 1%, 5%, and 10% levels, respectively.

Source: World Bank's PovcalNet database and IMF. Coverage: 2004-2018 period.

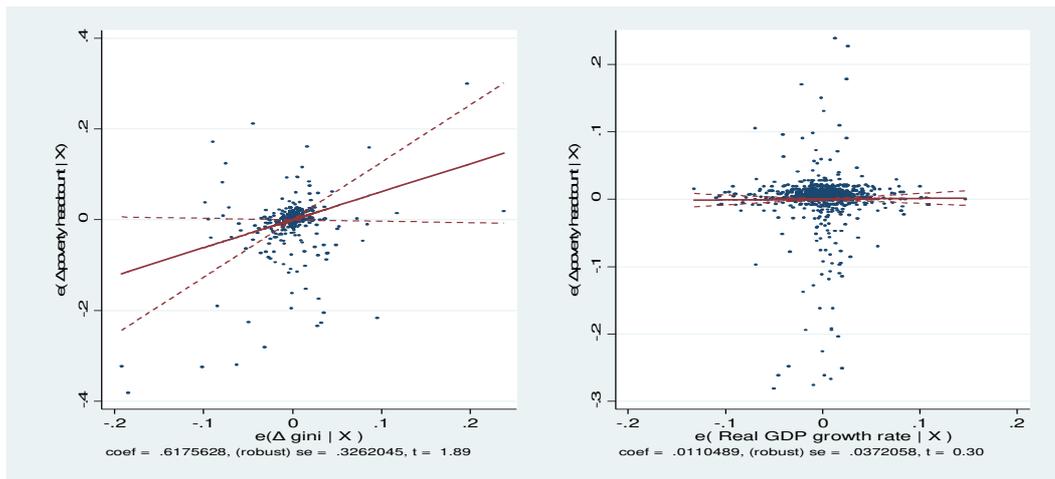

Figure 3: Marginal effects of growth, change in inequality on changes on poverty.

In Figure 3, we plot the marginal effects of all the regression coefficients shown in column 1. These marginal effects illustrate the limited effect of growth on poverty reduction, but a stronger effect of inequality on poverty.

*4.1. The role of inclusive financial development*

Table 3 presents the results of the role of financial inclusion on the relationship between inequality and poverty. We separately regress each of the four poverty measures on the financial inclusion index and its associated dimensions, totalling 12 regressions. In column 1, we use the financial inclusion index while using the financial outreach and usage dimensions in columns 2 and 3, respectively. In doing so, we can determine the role that bank branch outreach and the usage of financial services play on the link between income inequality and poverty.

The result in column 1 shows that neither the changes in the financial inclusion index nor its associated dimensions have any statistically significant direct effect on the changes in poverty. However, there is a negative and statistically significant interaction effect between all the measures of financial inclusion and Gini, except for column 2. These results suggest that the detrimental effect of inequality on poverty is strongly dampened in countries that have a greater level of financial inclusion. The interactions are all negative and of sizeable magnitude, particularly for financial outreach, suggesting that this dimension contributes the most to reduce the damaging effect of inequality on poverty.

Table 3. *Growth-redistribution decomposition of poverty: the role of financial inclusion.*

| Dependent variable | ΔHeadcount | | | ΔPoverty gap | | | ΔPoverty gap squared | | | ΔWatts | | |
|---|---|---|---|---|---|---|---|---|---|---|---|---|
| Financial inclusion type→ | ΔFinancial inclusion index | ΔFinancial outreach | ΔUsage | ΔFinancial inclusion index | ΔFinancial outreach | ΔUsage | ΔFinancial inclusion index | ΔFinancial outreach | ΔUsage | ΔFinancial inclusion index | ΔFinancial outreach | ΔUsage |
| | 1 | 2 | 3 | 4 | 5 | 6 | 7 | 8 | 9 | 10 | 11 | 12 |
| ΔGini | 0.789*** | 0.785** | 0.707** | 0.591*** | 0.600*** | 0.544*** | 0.441*** | 0.452*** | 0.408*** | 1.205*** | 1.234*** | 1.115*** |
| | [0.290] | [0.320] | [0.291] | [0.160] | [0.175] | [0.163] | [0.114] | [0.121] | [0.116] | [0.322] | [0.342] | [0.326] |
| GDP growth rate | 0.005 | 0.023 | 0.001 | -0.006 | 0.001 | -0.008 | -0.007 | -0.003 | -0.008 | -0.015 | -0.004 | -0.017 |
| | [0.041] | [0.041] | [0.039] | [0.023] | [0.023] | [0.022] | [0.015] | [0.015] | [0.015] | [0.041] | [0.042] | [0.041] |
| ΔFinancial inclusion | 0.073 | -0.12 | 0.092 | 0.029 | -0.034 | 0.03 | 0.017 | -0.012 | 0.014 | 0.045 | -0.043 | 0.041 |
| | [0.119] | [0.178] | [0.055] | [0.047] | [0.077] | [0.021] | [0.026] | [0.044] | [0.012] | [0.077] | [0.130] | [0.034] |
| ΔGini x ΔFinancial inclusion | -32.270** | -28.232 | -18.607* | -16.926** | -16.771* | -8.900** | -10.824** | -11.714** | -5.256* | -30.367** | -32.594* | -14.874* |
| | [16.012] | [20.257] | [10.358] | [6.520] | [9.044] | [4.187] | [4.254] | [5.770] | [2.689] | [12.053] | [16.533] | [7.558] |
| Constant | -0.007*** | -0.007*** | -0.007*** | -0.003** | -0.002** | -0.003** | -0.001* | -0.001* | -0.001* | -0.004** | -0.004** | -0.004* |
| | [0.002] | [0.002] | [0.002] | [0.001] | [0.001] | [0.001] | [0.001] | [0.001] | [0.001] | [0.002] | [0.002] | [0.002] |
| Observations | 856 | 856 | 856 | 856 | 856 | 856 | 856 | 856 | 856 | 856 | 856 | 856 |
| Adjusted $R^2$ | 0.153 | 0.136 | 0.142 | 0.245 | 0.237 | 0.228 | 0.263 | 0.263 | 0.246 | 0.27 | 0.269 | 0.252 |
| Number of country | 77 | 77 | 77 | 77 | 77 | 77 | 77 | 77 | 77 | 77 | 77 | 77 |
| Country fixed effects | Yes | Yes | Yes | Yes | Yes | Yes | Yes | Yes | Yes | Yes | Yes | Yes |
| Robust standard error clustered | Country | Country | Country | Country | Country | Country | Country | Country | Country | Country | Country | Country |

The dependent variables are changes in *headcount*, *poverty gap*, *poverty gap squared*, and *Watts* index. *ΔGini* is the change in income inequality. The *Financial Inclusion index* is a composite index, constructed based on two dimensions, namely *financial outreach* and *usage* dimensions: they enter the regression separately in changes. ***, **, and * indicate statistical significance at the 1%, 5%, and 10% levels, respectively. Source: World Bank's PovcalNet database and IMF. Coverage: 2004-2018 period.

In Figure 4, we plot the marginal effects of all the regression coefficients shown in column 1. These marginal effects illustrate the limited effect of growth on poverty reduction and the strong effect of financial inclusion via reducing the detrimental effect of inequality. Note that the interaction between Gini and financial outreach is not statistically significant, but it is significant with usage dimension when ΔHeadcount is the dependent variable. It suggests the importance of using financial services in reducing (headcount) poverty in a country where there is more inequality.

We also estimated alternative specifications where we interacted the financial inclusion index with GDP growth rate. These alternative interactions turned to be equal to zero and statistically insignificant, reason why we do not report them but are available upon request. These findings suggest financial inclusion helps poverty reduction via dampening the damaging effect of inequality but not though boosting the effect of growth.

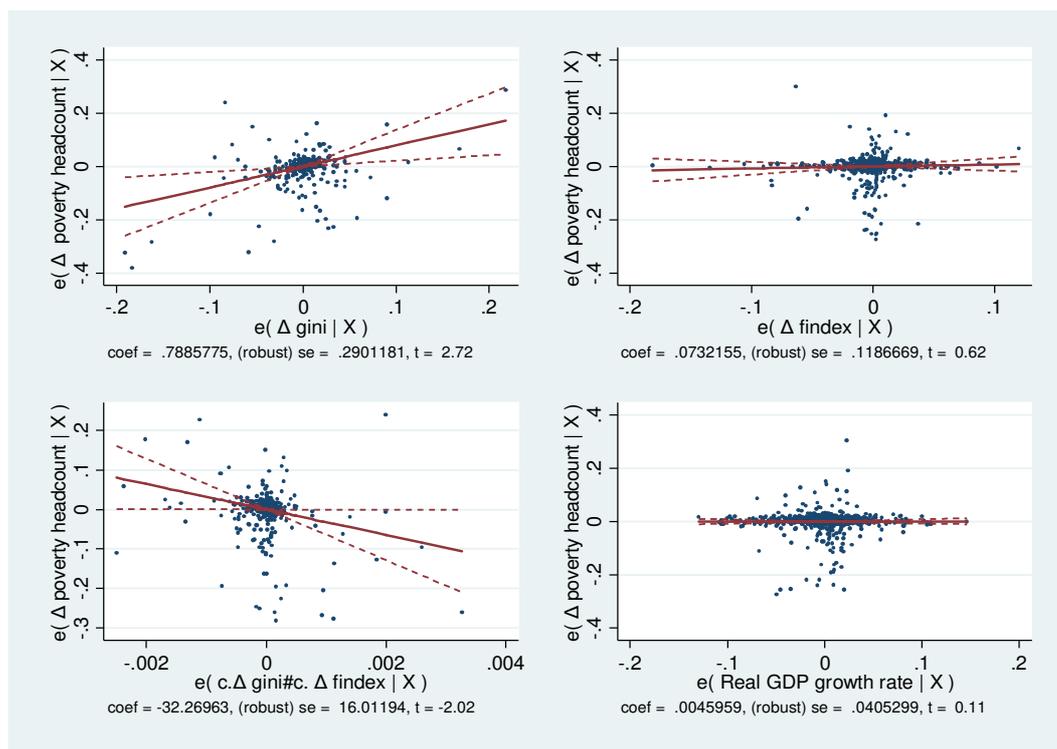

Figure 4: Marginal effect of growth, change in inequality, change on financial inclusion on poverty.

As a robustness check, we have re-estimated all our results using all variables in levels instead of changes. The results are reported in Table A.4, in the Appendix, which confirm our earlier findings, that greater inclusive financial development plays a positive role in reducing poverty in a country with higher inequality. The results are identical when we use financial outreach and usages dimensions, along with alternative measures of poverty.

*4.2. Alternative measure of financial inclusion*

As an additional robustness check, we use an alternative measure of financial inclusion, this time using demand-side measures. To this end, we collect data on account ownership at a financial institution for the year 2011, 2014 and 2017 from the Global Findex database. In addition, we collect information on this indicator for male and female separately. In Table 4, we replace our measures of financial inclusion with information on account ownership, as an alternative indicator of financial inclusion in eq. (3). As the Global Findex database starts reporting this information from the year 2011, we restrict our analysis for the period 2011-2018.[8] We use account ownership in the columns 1-4, while using the male and female variant of the indicator in columns 5-8 and 9-12, respectively. The results are consistent, as discussed above, using financial inclusion. It shows that the detrimental effect of inequality on poverty is dampened in countries that have higher account ownership at a financial institution. Furthermore, using other poverty measures (changes in the poverty gap, poverty gap squared, watts index) also provide similar results throughout. Moreover, taking columns 5-8, we also find similar results that higher account ownership by men reduces poverty in a country where there is a higher level of income inequality. Likewise, our female account ownership indicator also provides the same results. However, the magnitude of the coefficients is larger, implying that poverty reduction effect is greater for women compared to men where the former have more account ownership.

---

[8] For any gap year on account ownership, we use values from earlier waves of the survey. For instance, for 2012 and 2013, we use data for 2011.

Table 4. *Growth-redistribution decomposition of poverty: the role of financial inclusion (using an alternative measure of financial inclusion from Global Findex)*

| Dependent variable→ | ΔHeadcount | ΔPoverty gap | ΔPoverty gap squared | ΔWatts | ΔHeadcount | ΔPoverty gap | ΔPoverty gap squared | ΔWatts | ΔHeadcount | ΔPoverty gap | ΔPoverty gap squared | ΔWatts |
|---|---|---|---|---|---|---|---|---|---|---|---|---|
| Global findex→ | Account ownership at a financial institution | | | | Account ownership at a financial institution (Male) | | | | Account ownership at a financial institution (Female) | | | |
| | 1 | 2 | 3 | 4 | 5 | 6 | 7 | 8 | 9 | 10 | 11 | 12 |
| ΔGini | 0.922** | 0.694*** | 0.523*** | 1.430*** | 0.930** | 0.698** | 0.526** | 1.439** | 0.891** | 0.677*** | 0.511*** | 1.395*** |
| | [0.420] | [0.253] | [0.180] | [0.506] | [0.458] | [0.278] | [0.198] | [0.556] | [0.378] | [0.225] | [0.161] | [0.449] |
| GDP growth rate | 0.07 | 0.036 | 0.023 | 0.064 | 0.064 | 0.032 | 0.021 | 0.057 | 0.075 | 0.039 | 0.025 | 0.069 |
| | [0.071] | [0.042] | [0.029] | [0.080] | [0.071] | [0.042] | [0.029] | [0.080] | [0.070] | [0.042] | [0.029] | [0.080] |
| ΔAccount ownership | -0.007 | -0.015 | -0.015 | -0.039 | -0.007 | -0.016 | -0.015 | -0.04 | -0.004 | -0.012 | -0.013 | -0.033 |
| | [0.032] | [0.028] | [0.025] | [0.065] | [0.029] | [0.028] | [0.025] | [0.065] | [0.032] | [0.026] | [0.022] | [0.059] |
| ΔGini x ΔAccount ownership | -9.433** | -5.414** | -3.891** | -11.059** | -7.455* | -4.230** | -3.074** | -8.730** | -10.783*** | -6.238*** | -4.447*** | -12.649*** |
| | [4.203] | [2.215] | [1.536] | [4.401] | [4.079] | [2.051] | [1.392] | [3.962] | [3.930] | [2.097] | [1.463] | [4.221] |
| Constant | -0.010*** | -0.005** | -0.003* | -0.008* | -0.009*** | -0.005** | -0.003* | -0.008* | -0.010*** | -0.005** | -0.003** | -0.009** |
| | [0.004] | [0.002] | [0.002] | [0.004] | [0.004] | [0.002] | [0.002] | [0.004] | [0.003] | [0.002] | [0.002] | [0.004] |
| Observations | 391 | 391 | 391 | 391 | 391 | 391 | 391 | 391 | 391 | 391 | 391 | 391 |
| Adjusted $R^2$ | 0.208 | 0.283 | 0.291 | 0.299 | 0.188 | 0.266 | 0.276 | 0.282 | 0.227 | 0.3 | 0.306 | 0.315 |
| Number of country | 52 | 52 | 52 | 52 | 52 | 52 | 52 | 52 | 52 | 52 | 52 | 52 |
| Country fixed effects | Yes | Yes | Yes | Yes | Yes | Yes | Yes | Yes | Yes | Yes | Yes | Yes |
| Robust standard error clustered | Yes | Yes | Yes | Yes | Yes | Yes | Yes | Yes | Yes | Yes | Yes | Yes |

The dependent variables are changes in *headcount*, *poverty gap*, *poverty gap squared*, and *Watts* index. *ΔGini* is the change in income inequality. *Account* ownership is an indicator that refers to account ownership at a financial institution with male and female variants: they enter the regression separately in changes. ***, **, and * indicate statistical significance at the 1%, 5%, and 10% levels respectively. Source: World Bank's PovcalNet database and Global Findex. Coverage: 2011-2018 period.

*4.3. What are the main constraints for financial inclusion?*

The micro-data from the latest 2017 Global Findex survey gathered by the World Bank, also help us to understand what are the most important barriers to financial inclusion. For instance, Table 5 shows that there are significant differences in the main source of household emergency funds. The poorest households (in the bottom quintiles) rely more on family and friends than wealthier households (in top quintiles) who rely more on savings. Table 6 also shows that these differences exist between males and females. Women rely more on friends and family than men for emergency funds (37% versus 27%). Moreover, money from working is the main source of emergency for only 19% of women compared to 30% of men.

In terms of COVID-responses in Table 5 (Panels B and C) we learn that the majority of the population (about 60%) has received government transfers into one of their accounts, but surprisingly less than 5% of households have received a government transfer through a mobile phone. The social distancing imposed around the globe suggests that government transfers, particularly to poorest households, need to use more digital technologies such as mobile banking.

Table 5 also shows the main barriers preventing people from having a financial account. By far, the biggest constraint is lack of money (60% of households) across the board, regardless of household income (Panel D). Also, roughly 30% of households state that the main reason for them not to have an account is financial institution fees (Panel E). About 21% say that the main reason for not having an account is that the financial institution is too far away (Panel F). A similar percentage also claim that the main constraint in opening an account is not having the documentation required (Panel G).[9] These constraints prevent millions of households from benefiting from financial inclusion and limit countries' ability to make a significant dent in poverty.

---

[9] Table 5, panel H, shows that only a very small percentage (less than 7%) of respondents suggest that religion is one of the main constraints in opening an account.

Table 5. *Usage of emergency funds and constraints for financial access by quintiles*

| | Poorest | | | | Wealthiest |
|---|---|---|---|---|---|
| Quintiles: | I | II | III | IV | V |
| Panel A: Main source of emergency funds | | | Percent | | |
| Main source: Savings | 26.05 | 28.70 | 30.99 | 32.95 | 35.96 |
| Main source: Family or friends | 39.56 | 36.27 | 33.30 | 29.81 | 25.07 |
| Main source: Money from working | 19.43 | 21.78 | 24.06 | 26.11 | 29.78 |
| source: Borrowing from a bank, emp | 6.74 | 6.36 | 5.72 | 5.60 | 4.45 |
| Main source: Selling assets | 4.57 | 4.01 | 3.42 | 3.01 | 2.17 |
| Main source: Some other source | 2.09 | 1.83 | 1.71 | 1.69 | 1.93 |
| Panel B: If received government transfers into an account | | | | | |
| Yes | 56.41 | 58.55 | 61.04 | 63.80 | 65.34 |
| No | 42.56 | 40.63 | 38.15 | 35.61 | 33.97 |
| Panel C: If received government transfers through a mobile phone | | | | | |
| Yes | 4.07 | 3.41 | 4.44 | 4.51 | 5.34 |
| No | 95.07 | 95.72 | 94.78 | 94.63 | 93.82 |
| Panel D: Main reason for not having an account is lack of money | | | | | |
| Yes | 67.08 | 66.55 | 64.62 | 61.63 | 57.56 |
| No | 30.03 | 31.20 | 33.12 | 36.13 | 39.82 |
| Panel E: Main reason for not having an account is because it is too expensive | | | | | |
| Yes | 32.05 | 31.30 | 29.54 | 27.28 | 26.78 |
| No | 59.99 | 61.65 | 63.85 | 65.79 | 66.83 |
| Panel F: Main reason for not having an account is because institution is too far away | | | | | |
| Yes | 24.56 | 21.89 | 21.21 | 18.69 | 17.37 |
| No | 70.93 | 74.22 | 75.40 | 77.85 | 78.73 |
| Panel G: Main reason for not having an account is because lacks documentation | | | | | |
| Yes | 22.89 | 23.33 | 23.13 | 22.20 | 22.86 |
| No | 72.50 | 73.52 | 73.63 | 74.56 | 73.82 |
| Panel H: Main reason for not having an account is because of religious reasons | | | | | |
| Yes | 7.50 | 6.85 | 6.90 | 6.43 | 7.10 |
| No | 88.20 | 89.55 | 89.42 | 90.20 | 89.38 |

Source: Own estimates using World Bank's 2017 Global Findex.

Table 6. *Usage of emergency funds by males and females*

| | Male | Female |
|---|---|---|
| Main source of emergency funds | Percent | |
| Main source: Savings | 30.87 | 32.66 |
| Main source: Family or friends | 27.27 | 36.61 |
| Main source: Money from working | 30.26 | 19.15 |
| source: Borrowing from a bank, emp | 5.47 | 5.75 |
| Main source: Selling assets | 3.52 | 2.94 |
| Main source: Some other source | 1.77 | 1.90 |

Source: Own estimates using World Bank's 2017 Global Findex.

## 5. FORECASTS

In this section, we forecast the short-run effects of the likely changes in poverty due to the COVID-19 crisis. We use again the PovcalNet dataset from the year 2004 until 2018, using only the $1.90 a day poverty line. We use eq. (3) as the basis for our forecasts, as well as the overall index of financial inclusion we constructed earlier using the IMF data (IMF, 2020).

We explore three forecast scenarios. The first one assumes the forecasted reduction in GDP growth for each country as estimated by the IMF for the year 2020 and 2021 (IMF, 2020). The second scenario, on top of the fall in GDP growth also assumes that the Gini indices will increase by 1% in each country. The third scenario assumes that on top of the fall in GDP growth, there is an improvement of 10% in the financial inclusion index, assuming no changes in the Gini indices.

We produce these forecasts for *all* countries in Figure 5. In Figure 6, we show the forecasts only for the 78 low- and lower-middle-income countries analysed in this paper. Overall our forecasts suggest that poverty headcount could increase from 8% up to 14% across the three forecast scenarios, pushing over 400 million people into poverty by 2021. Poverty rises could be curbed with significant improvements in financial inclusion, as depicted by our third scenario in both Figures 5 and 6.

Our poverty forecasts are in the same range of other studies also using PovcalNet (Sumner, Hoy, & Ortiz-Juarez, 2020). We acknowledge that these forecasts are highly speculative as over 181 countries have also implemented recently urgent welfare and labour packages which we have not considered in our forecast directly (Gentilini et al., 2020). Nonetheless, the main take away from our forecast analysis is that improvements in financial inclusion could reduce the impact on poverty.

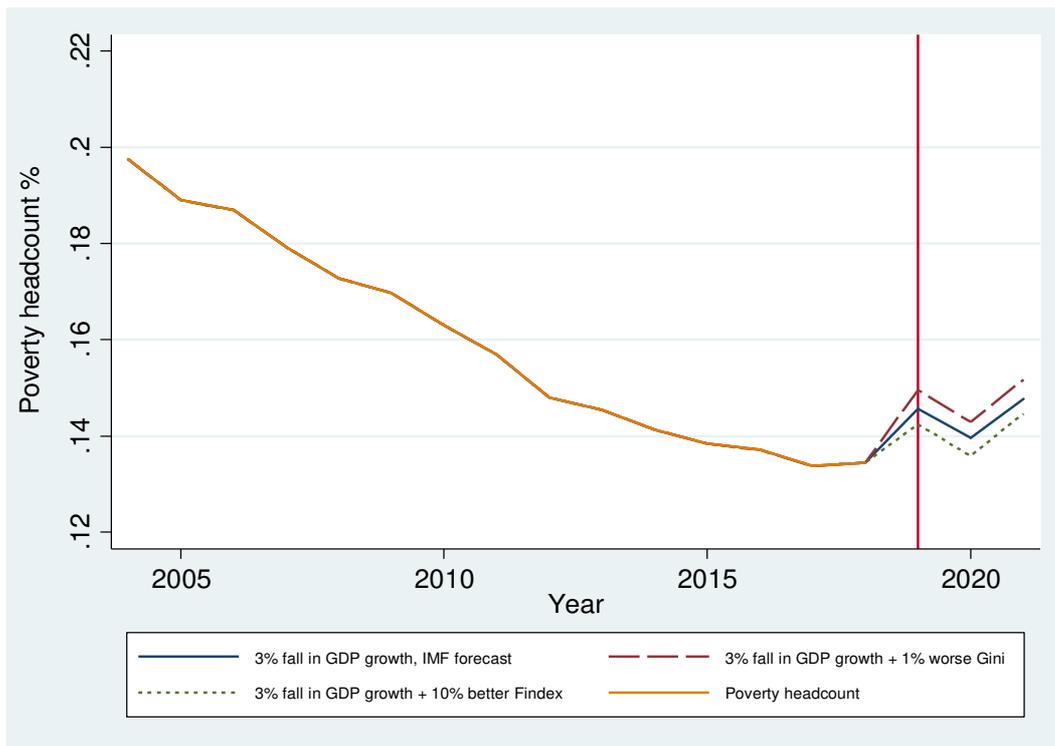

Figure 5. *Forecast poverty headcount $1.90 a day, all countries.*

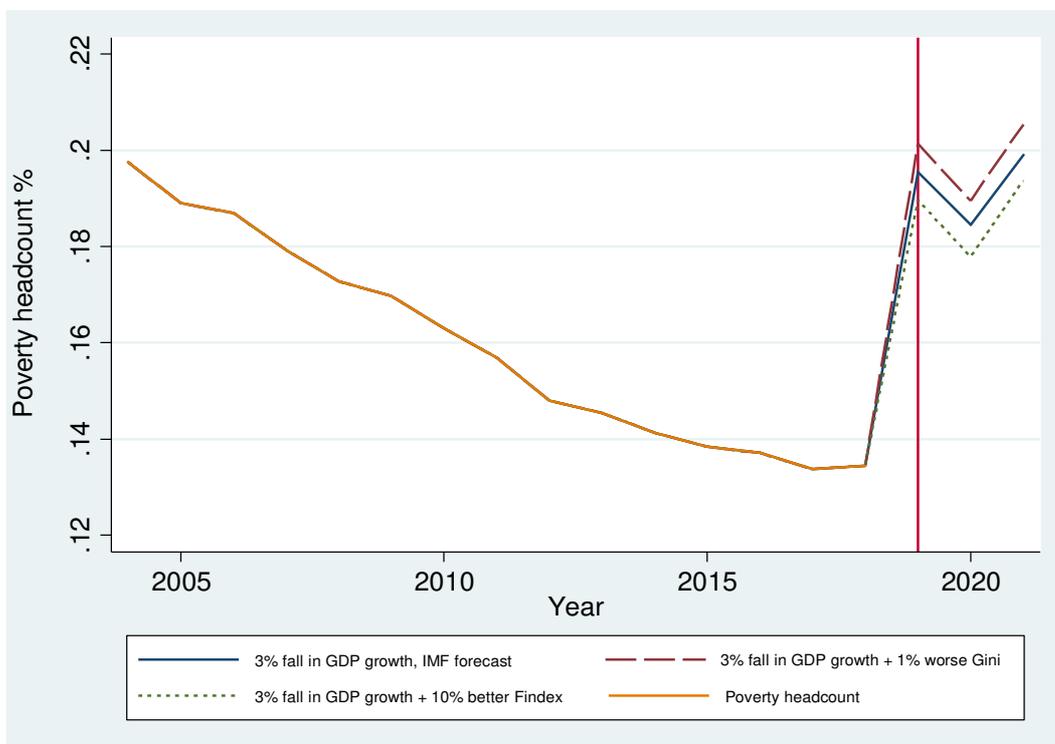

Figure 6. *Forecast poverty headcount $1.90 a day, including low- and lower-middle-income countries only.*

# 6. CONCLUSION

Using a sample of 78 low- and lower-middle-income countries, we showed that financial inclusion is instrumental in offsetting increases in poverty triggered by widening inequality. Furthermore, our forecast analysis shows that nearly 400 million people could be pushed into poverty, wiping out years of progress in poverty reduction if not adequate measures are taken. Our analysis suggests that financial inclusion could help curb some of the increases in poverty. Financial inclusion can also be a key tool to deliver urgent assistance, break inequality barriers that prevent millions of households from investing in productive investments, smoothing their consumption, and weathering the pandemic storm.

Our findings offer an important contribution to the literature on financial inclusion and poverty. Unlike other recent cross-country studies, we have shown that financial inclusion has an important role in poverty reduction, even in low- and lower-middle-income countries (e.g. Goksu et al., 2017; Park & Mercado, 2018). The effect of financial inclusion on poverty reduction is, however indirect, by helping to mitigate the detrimental impact that inequality has on poverty. Our findings are in line with the experimental literature which has found that financial inclusion helps to break inequality barriers among the most vulnerable groups (e.g. Banerjee, Karlan, & Zinman, 2015; Dupas & Robinson, 2013; Koomson et al., 2020; Li, 2018).

Financial inclusion has been regarded as a key complementary tool to address the Sustainable Development Goals (Chibba, 2009). The developing world has never before needed higher speed in accelerating financial inclusion as a complementary tool to reduce poverty. COVID-19 crises will require flexibility from all sectors involved, particularly from the financial industry. As millions of poor people and small firms have lost their livelihoods suddenly, they will need rapid access to government assistance and financial services such as savings and credit instruments. Financial institutions will have a crucial role to play to keep the economy afloat, contain potential regional contagion of financial collapse and help resuscitate small businesses once social distancing measures are eased out.

APPENDIX

Table A.1. *Financial inclusion index*

| Country | Income Level | 2004 | 2005 | 2006 | 2007 | 2008 | 2009 | 2010 | 2011 | 2012 | 2013 | 2014 | 2015 | 2016 | 2017 | 2018 |
|---|---|---|---|---|---|---|---|---|---|---|---|---|---|---|---|---|
| Afghanistan | Low income | 0.019 | 0.020 | 0.014 | 0.017 | 0.023 | 0.018 | 0.019 | 0.017 | 0.024 | 0.021 | 0.026 | 0.027 | 0.027 | 0.026 | 0.027 |
| Albania | Upper-middle income | 0.257 | 0.242 | 0.248 | 0.315 | 0.316 | 0.317 | 0.359 | 0.321 | 0.362 | 0.359 | 0.353 | 0.323 | 0.351 | 0.312 | 0.298 |
| Algeria | Upper-middle income | 0.085 | 0.086 | 0.091 | 0.098 | 0.099 | 0.071 | 0.079 | 0.082 | 0.086 | 0.088 | 0.090 | 0.090 | 0.089 | 0.089 | 0.090 |
| Angola | Lower-middle income | 0.007 | 0.008 | 0.013 | 0.016 | 0.020 | 0.024 | 0.032 | 0.037 | 0.042 | 0.047 | 0.049 | 0.050 | 0.050 | 0.051 | 0.048 |
| Argentina | Upper-middle income | 0.128 | 0.142 | 0.150 | 0.164 | 0.179 | 0.190 | 0.204 | 0.224 | 0.252 | 0.266 | 0.278 | 0.306 | 0.333 | 0.358 | 0.381 |
| Armenia | Lower-middle income | 0.095 | 0.133 | 0.144 | 0.161 | 0.181 | 0.185 | 0.208 | 0.247 | 0.285 | 0.317 | 0.362 | 0.379 | 0.396 | 0.402 | 0.430 |
| Austria | High income | 0.389 | 0.376 | 0.377 | 0.379 | 0.381 | 0.378 | 0.374 | 0.387 | 0.388 | 0.392 | 0.428 | 0.420 | 0.402 | 0.390 | 0.397 |
| Azerbaijan | Upper-middle income | 0.337 | 0.099 | 0.143 | 0.097 | 0.105 | 0.111 | 0.129 | 0.148 | 0.177 | 0.228 | 0.308 | 0.340 | 0.317 | 0.318 | 0.344 |
| Bahamas | High income | 0.471 | 0.458 | 0.479 | 0.523 | 0.522 | 0.506 | 0.473 | 0.479 | 0.462 | 0.482 | 0.478 | 0.458 | 0.476 | 0.460 | 0.461 |
| Bangladesh | Lower-middle income | 0.138 | 0.141 | 0.145 | 0.148 | 0.154 | 0.162 | 0.181 | 0.202 | 0.213 | 0.229 | 0.242 | 0.261 | 0.277 | 0.289 | 0.304 |
| Belgium | High income | 0.964 | 0.964 | 0.963 | 0.967 | 0.964 | 0.959 | 0.956 | 0.954 | 0.952 | 0.949 | 0.942 | 0.938 | 0.931 | 0.926 | 0.964 |
| Belize | Upper-middle income | 0.200 | 0.211 | 0.230 | 0.217 | 0.228 | 0.242 | 0.210 | 0.242 | 0.233 | 0.234 | 0.226 | 0.209 | 0.211 | 0.199 | 0.207 |
| Bhutan | Lower-middle income | 0.093 | 0.086 | 0.087 | 0.091 | 0.097 | 0.105 | 0.136 | 0.176 | 0.095 | 0.187 | 0.229 | 0.251 | 0.278 | 0.261 | 0.187 |
| Bolivia | Lower-middle income | 0.056 | 0.075 | 0.045 | 0.072 | 0.079 | 0.091 | 0.105 | 0.125 | 0.138 | 0.157 | 0.177 | 0.195 | 0.215 | 0.236 | 0.243 |
| Bosnia and Herzegovina | Upper-middle income | 0.241 | 0.276 | 0.269 | 0.287 | 0.314 | 0.333 | 0.317 | 0.336 | 0.345 | 0.364 | 0.370 | 0.367 | 0.362 | 0.365 | 0.372 |
| Botswana | Upper-middle income | 0.164 | 0.164 | 0.181 | 0.185 | 0.183 | 0.195 | 0.197 | 0.177 | 0.181 | 0.183 | 0.185 | 0.175 | 0.175 | 0.177 | 0.211 |
| Brazil | Upper-middle income | 0.550 | 0.576 | 0.560 | 0.302 | 0.323 | 0.344 | 0.367 | 0.391 | 0.537 | 0.573 | 0.572 | 0.556 | 0.599 | 0.602 | 0.576 |
| Brunei Darussalam | High income | 0.416 | 0.415 | 0.446 | 0.459 | 0.482 | 0.508 | 0.496 | 0.492 | 0.576 | 0.543 | 0.491 | 0.492 | 0.448 | 0.432 | 0.435 |
| Bulgaria | Upper-middle income | 0.516 | 0.452 | 0.503 | 0.552 | 0.600 | 0.613 | 0.607 | 0.587 | 0.589 | 0.582 | 0.575 | 0.564 | 0.546 | 0.548 | 0.550 |
| Burundi | Low income | 0.008 | 0.009 | 0.010 | 0.010 | 0.011 | 0.013 | 0.016 | 0.018 | 0.022 | 0.025 | 0.026 | 0.026 | 0.026 | 0.025 | 0.010 |
| Cambodia | Lower-middle income | 0.103 | 0.026 | 0.045 | 0.059 | 0.025 | 0.032 | 0.035 | 0.039 | 0.044 | 0.050 | 0.062 | 0.072 | 0.082 | 0.091 | 0.103 |
| Cameroon | Lower-middle income | 0.000 | 0.002 | 0.005 | 0.009 | 0.011 | 0.013 | 0.016 | 0.018 | 0.020 | 0.022 | 0.020 | 0.024 | 0.025 | 0.025 | 0.033 |
| Central African Republic | Low income | 0.003 | 0.000 | 0.000 | 0.002 | 0.003 | 0.005 | 0.006 | 0.008 | 0.008 | 0.011 | 0.009 | 0.009 | 0.011 | 0.010 | 0.009 |
| Chad | Low income | 0.005 | 0.005 | 0.005 | 0.004 | 0.005 | 0.005 | 0.006 | 0.006 | 0.006 | 0.006 | 0.007 | 0.007 | 0.008 | 0.008 | 0.007 |
| Chile | High income | 0.297 | 0.315 | 0.355 | 0.387 | 0.416 | 0.432 | 0.456 | 0.485 | 0.505 | 0.515 | 0.511 | 0.513 | 0.515 | 0.520 | 0.526 |
| China | Upper-middle income | 0.059 | 0.102 | 0.055 | 0.059 | 0.062 | 0.074 | 0.083 | 0.093 | 0.102 | 0.119 | 0.136 | 0.179 | 0.189 | 0.195 | 0.219 |
| Colombia | Upper-middle income | 0.214 | 0.228 | 0.260 | 0.241 | 0.243 | 0.259 | 0.255 | 0.271 | 0.288 | 0.304 | 0.321 | 0.337 | 0.341 | 0.353 | 0.363 |
| Comoros | | 0.019 | 0.011 | 0.015 | 0.023 | 0.019 | 0.024 | 0.018 | 0.039 | 0.039 | 0.042 | 0.044 | 0.045 | 0.048 | 0.049 | 0.050 |
| Congo, Democratic Republic of | Low income | 0.005 | 0.003 | 0.003 | 0.002 | 0.002 | 0.002 | 0.003 | 0.003 | 0.006 | 0.008 | 0.008 | 0.005 | 0.003 | 0.002 | 0.006 |
| Congo, Republic of | Lower-middle income | 0.002 | 0.003 | 0.003 | 0.008 | 0.010 | 0.011 | 0.014 | 0.017 | 0.019 | 0.022 | 0.030 | 0.034 | 0.008 | 0.010 | 0.019 |
| Costa Rica | Upper-middle income | 0.309 | 0.301 | 0.296 | 0.329 | 0.338 | 0.340 | 0.341 | 0.372 | 0.380 | 0.419 | 0.440 | 0.429 | 0.480 | 0.509 | 0.526 |
| Croatia | Upper-middle income | 0.283 | 0.275 | 0.290 | 0.311 | 0.358 | 0.346 | 0.420 | 0.421 | 0.411 | 0.409 | 0.393 | 0.391 | 0.381 | 0.364 | 0.360 |
| Czech Republic | High income | 0.292 | 0.304 | 0.313 | 0.332 | 0.349 | 0.353 | 0.370 | 0.390 | 0.406 | 0.431 | 0.451 | 0.368 | 0.369 | 0.375 | 0.383 |
| Djibouti | Lower-middle income | 0.017 | 0.038 | 0.018 | 0.030 | 0.016 | 0.021 | 0.026 | 0.030 | 0.033 | 0.041 | 0.048 | 0.056 | 0.051 | 0.068 | 0.030 |
| Dominican Republic | Upper-middle income | 0.208 | 0.242 | 0.193 | 0.204 | 0.220 | 0.207 | 0.217 | 0.237 | 0.224 | 0.245 | 0.254 | 0.279 | 0.291 | 0.291 | 0.291 |
| Ecuador | Upper-middle income | 0.145 | 0.119 | 0.112 | 0.147 | 0.162 | 0.177 | 0.187 | 0.179 | 0.176 | 0.171 | 0.185 | 0.189 | 0.190 | 0.190 | 0.195 |
| Egypt, Arab Republic of | Lower-middle income | 0.076 | 0.071 | 0.072 | 0.075 | 0.078 | 0.082 | 0.076 | 0.076 | 0.080 | 0.091 | 0.094 | 0.098 | 0.105 | 0.129 | 0.139 |
| El Salvador | Lower-middle income | 0.245 | 0.228 | 0.260 | 0.251 | 0.261 | 0.258 | 0.253 | 0.262 | 0.264 | 0.278 | 0.279 | 0.285 | 0.299 | 0.326 | 0.331 |
| Equatorial Guinea | Upper-middle income | 0.016 | 0.017 | 0.019 | 0.023 | 0.025 | 0.027 | 0.029 | 0.034 | 0.038 | 0.037 | 0.041 | 0.047 | 0.048 | 0.016 | 0.023 |
| Estonia | High income | 0.542 | 0.500 | 0.564 | 0.628 | 0.584 | 0.573 | 0.593 | 0.575 | 0.485 | 0.477 | 0.471 | 0.449 | 0.442 | 0.440 | 0.453 |
| Eswatini | Lower-middle income | 0.118 | 0.126 | 0.129 | 0.114 | 0.136 | 0.136 | 0.141 | 0.128 | 0.145 | 0.151 | 0.169 | 0.173 | 0.161 | 0.159 | 0.164 |
| Fiji | Upper-middle income | 0.131 | 0.153 | 0.187 | 0.171 | 0.186 | 0.187 | 0.200 | 0.211 | 0.209 | 0.228 | 0.255 | 0.275 | 0.301 | 0.316 | 0.313 |
| Finland | High income | 0.325 | 0.285 | 0.211 | 0.340 | 0.208 | 0.205 | 0.339 | 0.336 | 0.332 | 0.323 | 0.319 | 0.315 | 0.295 | 0.162 | 0.254 |
| Gabon | Upper-middle income | 0.041 | 0.026 | 0.027 | 0.030 | 0.041 | 0.042 | 0.047 | 0.053 | 0.073 | 0.107 | 0.047 | 0.047 | 0.041 | 0.041 | 0.107 |
| Georgia | Lower-middle income | 0.072 | 0.102 | 0.141 | 0.192 | 0.257 | 0.258 | 0.282 | 0.326 | 0.385 | 0.427 | 0.478 | 0.482 | 0.521 | 0.572 | 0.579 |
| Ghana | Lower-middle income | 0.118 | 0.035 | 0.039 | 0.051 | 0.051 | 0.054 | 0.065 | 0.071 | 0.079 | 0.086 | 0.087 | 0.114 | 0.112 | 0.132 | 0.140 |
| Greece | High income | 0.693 | 0.679 | 0.696 | 0.712 | 0.727 | 0.725 | 0.719 | 0.707 | 0.698 | 0.662 | 0.642 | 0.636 | 0.629 | 0.626 | 0.620 |
| Guatemala | Lower-middle income | 0.206 | 0.210 | 0.212 | 0.267 | 0.288 | 0.307 | 0.323 | 0.348 | 0.368 | 0.395 | 0.409 | 0.413 | 0.404 | 0.390 | 0.345 |
| Guinea | Low income | 0.003 | 0.004 | 0.005 | 0.006 | 0.007 | 0.009 | 0.010 | 0.012 | 0.014 | 0.015 | 0.019 | 0.021 | 0.020 | 0.025 | 0.025 |
| Guyana | Upper-middle income | 0.120 | 0.125 | 0.105 | 0.124 | 0.125 | 0.145 | 0.149 | 0.155 | 0.159 | 0.163 | 0.162 | 0.160 | 0.157 | 0.156 | 0.153 |
| Haiti | Low income | 0.045 | 0.049 | 0.053 | 0.056 | 0.056 | 0.059 | 0.057 | 0.058 | 0.059 | 0.054 | 0.051 | 0.052 | 0.052 | 0.054 | 0.056 |
| Honduras | Lower-middle income | 0.148 | 0.150 | 0.164 | 0.187 | 0.194 | 0.197 | 0.186 | 0.196 | 0.214 | 0.218 | 0.229 | 0.235 | 0.240 | 0.239 | 0.241 |
| Hungary | High income | 0.290 | 0.296 | 0.298 | 0.321 | 0.358 | 0.362 | 0.363 | 0.364 | 0.358 | 0.357 | 0.348 | 0.339 | 0.341 | 0.346 | 0.342 |
| Iceland | High income | 0.697 | 0.703 | 0.697 | 0.697 | 0.692 | 0.668 | 0.670 | 0.664 | 0.651 | 0.643 | 0.638 | 0.623 | 0.637 | 0.635 | 0.624 |
| India | Lower-middle income | 0.201 | 0.139 | 0.143 | 0.152 | 0.165 | 0.180 | 0.198 | 0.215 | 0.237 | 0.264 | 0.308 | 0.344 | 0.378 | 0.403 | 0.413 |
| Indonesia | Lower-middle income | 0.100 | 0.105 | 0.104 | 0.109 | 0.118 | 0.128 | 0.138 | 0.178 | 0.228 | 0.260 | 0.279 | 0.289 | 0.305 | 0.361 | 0.362 |
| Ireland | High income | 0.429 | 0.422 | 0.424 | 0.476 | 0.485 | 0.480 | 0.464 | 0.438 | 0.419 | 0.405 | 0.478 | 0.483 | 0.482 | 0.450 | 0.429 |
| Italy | High income | 0.563 | 0.569 | 0.584 | 0.602 | 0.640 | 0.641 | 0.633 | 0.640 | 0.634 | 0.613 | 0.610 | 0.626 | 0.613 | 0.599 | 0.573 |
| Jamaica | Upper-middle income | 0.236 | 0.233 | 0.232 | 0.226 | 0.228 | 0.227 | 0.223 | 0.218 | 0.225 | 0.227 | 0.235 | 0.240 | 0.246 | 0.293 | 0.306 |
| Japan | High income | 0.957 | 0.957 | 0.956 | 0.955 | 0.955 | 0.955 | 0.955 | 0.955 | 0.956 | 0.955 | 0.955 | 0.956 | 0.956 | 0.956 | 0.956 |
| Jordan | Lower-middle income | 0.193 | 0.194 | 0.194 | 0.188 | 0.189 | 0.193 | 0.174 | 0.174 | 0.173 | 0.168 | 0.165 | 0.163 | 0.172 | 0.175 | 0.193 |
| Kenya | Lower-middle income | 0.044 | 0.030 | 0.037 | 0.061 | 0.063 | 0.071 | 0.093 | 0.099 | 0.106 | 0.133 | 0.163 | 0.193 | 0.217 | 0.231 | 0.253 |
| Kiribati | Lower-middle income | 0.072 | 0.062 | 0.062 | 0.062 | 0.072 | 0.072 | 0.072 | 0.060 | 0.062 | 0.072 | 0.062 | 0.062 | 0.062 | 0.062 | 0.062 |
| Korea, Republic of | High income | 0.867 | 0.862 | 0.878 | 0.883 | 0.888 | 0.885 | 0.886 | 0.888 | 0.890 | 0.888 | 0.882 | 0.879 | 0.875 | 0.869 | 0.869 |
| Kosobo | | 0.244 | 0.198 | 0.211 | 0.233 | 0.263 | 0.279 | 0.291 | 0.327 | 0.336 | 0.329 | 0.321 | 0.319 | 0.327 | 0.289 | 0.271 |
| Lao People's Democratic Republic | Lower-middle income | 0.086 | 0.066 | 0.086 | 0.082 | 0.066 | 0.074 | 0.082 | 0.096 | 0.077 | 0.073 | 0.086 | 0.097 | 0.102 | 0.110 | 0.116 |
| Latvia | High income | 0.333 | 0.363 | 0.402 | 0.505 | 0.535 | 0.528 | 0.531 | 0.542 | 0.515 | 0.500 | 0.435 | 0.413 | 0.398 | 0.412 | 0.403 |

Table A.1. *Financial inclusion index, cont.*

| Country | Income Level | 2004 | 2005 | 2006 | 2007 | 2008 | 2009 | 2010 | 2011 | 2012 | 2013 | 2014 | 2015 | 2016 | 2017 | 2018 |
|---|---|---|---|---|---|---|---|---|---|---|---|---|---|---|---|---|
| Lebanon | Upper-middle income | 0.456 | 0.405 | 0.418 | 0.436 | 0.463 | 0.486 | 0.504 | 0.507 | 0.515 | 0.500 | 0.490 | 0.491 | 0.491 | 0.501 | 0.512 |
| Lesotho | Lower-middle income | 0.063 | 0.050 | 0.071 | 0.065 | 0.052 | 0.057 | 0.065 | 0.070 | 0.064 | 0.075 | 0.076 | 0.081 | 0.085 | 0.087 | 0.092 |
| Liberia | Low income | 0.034 | 0.011 | 0.011 | 0.011 | 0.013 | 0.020 | 0.026 | 0.030 | 0.034 | 0.043 | 0.045 | 0.045 | 0.026 | 0.020 | 0.045 |
| Libyan Arab Jamahirya | Upper-middle income | 0.038 | 0.038 | 0.042 | 0.042 | 0.043 | 0.047 | 0.048 | 0.046 | 0.045 | 0.046 | 0.047 | 0.046 | 0.046 | 0.046 | 0.042 |
| Madagascar | Low income | 0.005 | 0.006 | 0.007 | 0.008 | 0.010 | 0.011 | 0.018 | 0.013 | 0.014 | 0.016 | 0.019 | 0.021 | 0.022 | 0.025 | 0.006 |
| Malawi | Low income | 0.031 | 0.033 | 0.033 | 0.040 | 0.039 | 0.047 | 0.043 | 0.041 | 0.046 | 0.048 | 0.049 | 0.056 | 0.033 | 0.043 | 0.041 |
| Malaysia | Upper-middle income | 0.377 | 0.398 | 0.402 | 0.418 | 0.427 | 0.441 | 0.450 | 0.457 | 0.467 | 0.488 | 0.484 | 0.464 | 0.457 | 0.453 | 0.451 |
| Maldives | Upper-middle income | 0.334 | 0.343 | 0.369 | 0.444 | 0.448 | 0.424 | 0.424 | 0.461 | 0.454 | 0.489 | 0.500 | 0.516 | 0.527 | 0.528 | 0.532 |
| Malta | High income | 0.824 | 0.831 | 0.835 | 0.872 | 0.888 | 0.899 | 0.901 | 0.909 | 0.905 | 0.904 | 0.898 | 0.898 | 0.891 | 0.885 | 0.876 |
| Mauritania | Lower-middle income | 0.032 | 0.054 | 0.054 | 0.032 | 0.040 | 0.045 | 0.039 | 0.033 | 0.034 | 0.043 | 0.044 | 0.048 | 0.054 | 0.056 | 0.045 |
| Mauritius | Upper-middle income | 0.531 | 0.567 | 0.569 | 0.604 | 0.621 | 0.648 | 0.661 | 0.682 | 0.691 | 0.701 | 0.690 | 0.686 | 0.661 | 0.655 | 0.638 |
| Mexico | Upper-middle income | 0.127 | 0.133 | 0.148 | 0.159 | 0.207 | 0.214 | 0.234 | 0.206 | 0.219 | 0.253 | 0.217 | 0.223 | 0.234 | 0.228 | 0.237 |
| Micronesia, Federated States of | Lower-middle income | 0.108 | 0.108 | 0.110 | 0.109 | 0.110 | 0.128 | 0.141 | 0.143 | 0.153 | 0.152 | 0.152 | 0.137 | 0.132 | 0.128 | 0.126 |
| Moldova | Lower-middle income | 0.237 | 0.260 | 0.278 | 0.310 | 0.345 | 0.351 | 0.365 | 0.386 | 0.397 | 0.416 | 0.421 | 0.289 | 0.309 | 0.320 | 0.330 |
| Mongolia | Lower-middle income | 0.226 | 0.174 | 0.213 | 0.321 | 0.246 | 0.254 | 0.273 | 0.300 | 0.327 | 0.353 | 0.377 | 0.401 | 0.463 | 0.433 | 0.464 |
| Montenegro | Upper-middle income | 0.432 | 0.356 | 0.359 | 0.308 | 0.457 | 0.452 | 0.447 | 0.460 | 0.468 | 0.483 | 0.468 | 0.473 | 0.459 | 0.462 | 0.511 |
| Morocco | Lower-middle income | 0.135 | 0.132 | 0.130 | 0.141 | 0.167 | 0.200 | 0.227 | 0.237 | 0.248 | 0.257 | 0.265 | 0.259 | 0.271 | 0.287 | 0.293 |
| Mozambique | Low income | 0.024 | 0.014 | 0.018 | 0.021 | 0.023 | 0.027 | 0.030 | 0.034 | 0.037 | 0.045 | 0.049 | 0.056 | 0.062 | 0.058 | 0.058 |
| Myanmar | Lower-middle income | 0.021 | 0.025 | 0.019 | 0.020 | 0.020 | 0.021 | 0.018 | 0.018 | 0.021 | 0.026 | 0.032 | 0.038 | 0.042 | 0.046 | 0.057 |
| Namibia | Upper-middle income | 0.081 | 0.119 | 0.099 | 0.105 | 0.158 | 0.181 | 0.192 | 0.202 | 0.198 | 0.213 | 0.238 | 0.258 | 0.293 | 0.238 | 0.081 |
| Nepal | Low income | 0.137 | 0.129 | 0.091 | 0.133 | 0.141 | 0.138 | 0.112 | 0.158 | 0.104 | 0.108 | 0.119 | 0.131 | 0.145 | 0.172 | 0.210 |
| Netherlands | High income | 0.694 | 0.691 | 0.703 | 0.709 | 0.706 | 0.708 | 0.692 | 0.660 | 0.627 | 0.610 | 0.620 | 0.574 | 0.549 | 0.526 | 0.518 |
| Nicaragua | Lower-middle income | 0.065 | 0.076 | 0.078 | 0.089 | 0.094 | 0.085 | 0.076 | 0.082 | 0.084 | 0.090 | 0.099 | 0.109 | 0.126 | 0.133 | 0.123 |
| North Macedonia | Upper-middle income | 0.472 | 0.325 | 0.412 | 0.322 | 0.406 | 0.438 | 0.452 | 0.444 | 0.452 | 0.467 | 0.481 | 0.498 | 0.500 | 0.503 | 0.505 |
| Norway | High income | 0.210 | 0.202 | 0.218 | 0.221 | 0.246 | 0.217 | 0.213 | 0.214 | 0.209 | 0.209 | 0.204 | 0.200 | 0.188 | 0.182 | 0.186 |
| Oman | High income | 0.167 | 0.166 | 0.177 | 0.185 | 0.196 | 0.200 | 0.205 | 0.211 | 0.204 | 0.215 | 0.203 | 0.202 | 0.201 | 0.197 | 0.196 |
| Pakistan | Lower-middle income | 0.054 | 0.056 | 0.064 | 0.069 | 0.073 | 0.073 | 0.076 | 0.081 | 0.085 | 0.091 | 0.098 | 0.105 | 0.111 | 0.117 | 0.120 |
| Panama | Upper-middle income | 0.255 | 0.383 | 0.253 | 0.252 | 0.253 | 0.255 | 0.278 | 0.295 | 0.318 | 0.346 | 0.367 | 0.383 | 0.391 | 0.392 | 0.393 |
| Papua New Guinea | Lower-middle income | 0.061 | 0.061 | 0.050 | 0.049 | 0.061 | 0.051 | 0.064 | 0.056 | 0.066 | 0.066 | 0.061 | 0.057 | 0.062 | 0.068 | 0.069 |
| Paraguay | Upper-middle income | 0.044 | 0.043 | 0.048 | 0.046 | 0.051 | 0.044 | 0.057 | 0.062 | 0.074 | 0.073 | 0.081 | 0.083 | 0.091 | 0.118 | |
| Peru | Upper-middle income | 0.113 | 0.083 | 0.095 | 0.114 | 0.141 | 0.143 | 0.161 | 0.182 | 0.196 | 0.212 | 0.246 | 0.331 | 0.337 | 0.337 | 0.355 |
| Philippines | Lower-middle income | 0.103 | 0.107 | 0.110 | 0.109 | 0.115 | 0.119 | 0.130 | 0.142 | 0.144 | 0.158 | 0.166 | 0.176 | 0.185 | 0.192 | 0.202 |
| Poland | High income | 0.481 | 0.501 | 0.492 | 0.529 | 0.537 | 0.565 | 0.550 | 0.552 | 0.551 | 0.549 | 0.565 | 0.578 | 0.592 | 0.596 | 0.614 |
| Portugal | High income | 0.779 | 0.790 | 0.799 | 0.818 | 0.837 | 0.841 | 0.846 | 0.841 | 0.797 | 0.793 | 0.785 | 0.718 | 0.719 | 0.698 | 0.690 |
| Romania | Upper-middle income | 0.176 | 0.170 | 0.199 | 0.197 | 0.259 | 0.261 | 0.260 | 0.266 | 0.258 | 0.249 | 0.246 | 0.244 | 0.238 | 0.233 | 0.229 |
| Rwanda | Low income | 0.002 | 0.005 | 0.005 | 0.008 | 0.048 | 0.055 | 0.063 | 0.074 | 0.082 | 0.086 | 0.087 | 0.084 | 0.082 | 0.086 | 0.101 |
| Samoa | Upper-middle income | 0.133 | 0.150 | 0.162 | 0.176 | 0.188 | 0.204 | 0.207 | 0.235 | 0.247 | 0.305 | 0.303 | 0.300 | 0.349 | 0.355 | 0.360 |
| San Marino | High income | 0.966 | 0.982 | 0.954 | 0.988 | 0.993 | 1.000 | 0.967 | 0.952 | 0.949 | 0.974 | 0.984 | 0.923 | 0.923 | 0.921 | 0.915 |
| Sao Tome and Principe | Lower-middle income | | | | | | | | 0.242 | 0.277 | 0.297 | 0.287 | 0.323 | 0.264 | 0.277 | 0.260 |
| Saudi Arabia | High income | 0.215 | 0.217 | 0.177 | 0.234 | 0.241 | 0.195 | 0.201 | 0.202 | 0.204 | 0.228 | 0.235 | 0.246 | 0.256 | 0.266 | 0.270 |
| Serbia | Upper-middle income | 0.174 | 0.159 | 0.181 | 0.218 | 0.247 | 0.246 | 0.241 | 0.236 | 0.273 | 0.257 | 0.246 | 0.252 | 0.261 | 0.248 | 0.251 |
| Seychelles | High income | 0.415 | 0.438 | 0.453 | 0.451 | 0.497 | 0.513 | 0.517 | 0.529 | 0.581 | 0.603 | 0.644 | 0.640 | 0.662 | 0.694 | 0.702 |
| Solomon Islands | Lower-middle income | 0.056 | 0.058 | 0.058 | 0.059 | 0.066 | 0.066 | 0.061 | 0.059 | 0.061 | 0.053 | 0.077 | 0.080 | 0.083 | 0.086 | 0.077 |
| South Africa | Upper-middle income | 0.136 | 0.150 | 0.172 | 0.183 | 0.203 | 0.232 | 0.248 | 0.273 | 0.302 | 0.322 | 0.339 | 0.335 | 0.340 | 0.344 | 0.316 |
| South Sudan | Low income | | | | | | 0.013 | 0.011 | 0.009 | 0.004 | 0.005 | 0.008 | 0.012 | 0.013 | 0.016 | 0.014 | 0.013 |
| Spain | High income | 0.726 | 0.744 | 0.752 | 0.772 | 0.773 | 0.776 | 0.771 | 0.752 | 0.734 | 0.724 | 0.711 | 0.697 | 0.677 | 0.690 | 0.677 |
| Suriname | Upper-middle income | 0.261 | 0.255 | 0.269 | 0.240 | 0.272 | 0.279 | 0.282 | 0.267 | 0.272 | 0.276 | 0.296 | 0.260 | 0.288 | 0.283 | 0.296 |
| Sweden | High income | 0.559 | 0.561 | 0.561 | 0.554 | 0.557 | 0.565 | 0.530 | 0.540 | 0.552 | 0.559 | 0.558 | 0.553 | 0.541 | 0.534 | 0.543 |
| Switzerland | High income | 0.776 | 0.779 | 0.775 | 0.777 | 0.771 | 0.787 | 0.787 | 0.796 | 0.791 | 0.811 | 0.808 | 0.774 | 0.723 | 0.710 | 0.693 |
| Syrian Arab Republic | Lower-middle income | 0.041 | 0.039 | 0.040 | 0.041 | 0.045 | 0.055 | 0.058 | 0.060 | 0.060 | 0.048 | 0.060 | 0.058 | 0.045 | 0.042 | 0.060 |
| Tanzania | Low income | 0.026 | 0.036 | 0.027 | 0.026 | 0.029 | 0.028 | 0.033 | 0.037 | 0.039 | 0.040 | 0.062 | 0.046 | 0.046 | 0.029 | 0.026 |
| Thailand | Upper-middle income | 0.206 | 0.238 | 0.275 | 0.301 | 0.336 | 0.353 | 0.375 | 0.385 | 0.412 | 0.434 | 0.456 | 0.465 | 0.468 | 0.480 | 0.487 |
| Tonga | Upper-middle income | 0.208 | 0.232 | 0.249 | 0.204 | 0.223 | 0.208 | 0.200 | 0.204 | 0.213 | 0.234 | 0.232 | 0.287 | 0.302 | 0.285 | 0.344 | 0.358 |
| Trinidad and Tobago | High income | | | | | | 0.370 | 0.383 | 0.395 | 0.395 | 0.401 | 0.410 | 0.413 | 0.413 | 0.414 | 0.404 | 0.401 |
| Turkey | Upper-middle income | 0.289 | 0.277 | 0.288 | 0.307 | 0.311 | 0.319 | 0.470 | 0.491 | 0.514 | 0.561 | 0.557 | 0.570 | 0.592 | 0.627 | 0.636 |
| Uganda | Low income | 0.017 | 0.019 | 0.021 | 0.024 | 0.030 | 0.034 | 0.037 | 0.036 | 0.041 | 0.042 | 0.044 | 0.049 | 0.051 | 0.063 | 0.076 |
| Ukraine | Lower-middle income | 0.435 | 0.420 | 0.460 | 0.517 | 0.545 | 0.566 | 0.573 | 0.581 | 0.592 | 0.486 | 0.479 | 0.453 | 0.447 | 0.479 | 0.421 |
| United Arab Emirates | High income | 0.244 | 0.277 | 0.270 | 0.270 | 0.283 | 0.300 | 0.306 | 0.315 | 0.320 | 0.315 | 0.335 | 0.358 | 0.367 | 0.374 | 0.366 |
| United Kingdom | High income | 0.450 | 0.446 | 0.437 | 0.437 | 0.436 | 0.433 | 0.428 | 0.424 | 0.413 | 0.430 | 0.450 | 0.450 | 0.450 | 0.450 | 0.421 |
| Uzbekistan | Lower-middle income | 0.165 | 0.185 | 0.191 | 0.196 | 0.203 | 0.205 | 0.215 | 0.237 | 0.241 | 0.261 | 0.264 | 0.239 | 0.259 | 0.319 | 0.317 |
| Vanuatu | Lower-middle income | 0.158 | 0.165 | 0.165 | 0.175 | 0.184 | 0.180 | 0.188 | 0.195 | 0.205 | 0.206 | 0.200 | 0.209 | 0.209 | 0.208 | 0.206 |
| Venezuela, Republica Bolivariana | Upper-middle income | 0.144 | 0.094 | 0.109 | 0.128 | 0.131 | 0.137 | 0.144 | 0.150 | 0.154 | 0.173 | 0.186 | 0.219 | 0.131 | 0.219 | 0.154 |
| Vietnam | Lower-middle income | 0.094 | 0.149 | 0.135 | 0.145 | 0.120 | 0.159 | 0.086 | 0.123 | 0.136 | 0.149 | 0.164 | 0.175 | 0.189 | 0.188 | 0.207 |
| West Bank and Gaza | Lower-middle income | 0.215 | 0.305 | 0.211 | 0.223 | 0.233 | 0.248 | 0.257 | 0.285 | 0.298 | 0.304 | 0.317 | 0.330 | 0.349 | 0.363 | 0.419 |
| Yemen, Republic of | Lower-middle income | 0.018 | 0.019 | 0.020 | 0.018 | 0.020 | 0.022 | 0.021 | 0.020 | 0.022 | 0.023 | 0.026 | 0.025 | 0.026 | 0.026 | 0.025 |
| Zambia | Lower-middle income | 0.043 | 0.043 | 0.043 | 0.047 | 0.048 | 0.050 | 0.033 | 0.036 | 0.043 | 0.053 | 0.059 | 0.061 | 0.062 | 0.063 | 0.057 |
| Zimbabwe | Low income | 0.071 | 0.076 | 0.076 | 0.044 | 0.042 | 0.032 | 0.039 | 0.038 | 0.064 | 0.071 | 0.066 | 0.042 | 0.043 | 0.078 | 0.106 |

Source: Own estimates using Financial Access Survey, 2004-2018 (IMF, 2019). A few countries are missing in these tables as some of their information was not available at the time of writing this paper.

Table A.2. *Financial outreach index*

| Country | 2004 | 2005 | 2006 | 2007 | 2008 | 2009 | 2010 | 2011 | 2012 | 2013 | 2014 | 2015 | 2016 | 2017 | 2018 |
|---|---|---|---|---|---|---|---|---|---|---|---|---|---|---|---|
| Afghanistan | 0.001 | 0.002 | 0.004 | 0.005 | 0.007 | 0.011 | 0.012 | 0.012 | 0.011 | 0.012 | 0.013 | 0.012 | 0.012 | 0.012 | 0.015 |
| Albania | 0.076 | 0.105 | 0.132 | 0.181 | 0.241 | 0.257 | 0.262 | 0.266 | 0.267 | 0.263 | 0.251 | 0.252 | 0.247 | 0.234 | 0.223 |
| Algeria | 0.024 | 0.025 | 0.028 | 0.031 | 0.033 | 0.035 | 0.036 | 0.037 | 0.037 | 0.038 | 0.040 | 0.042 | 0.042 | 0.043 | 0.043 |
| Angola | 0.009 | 0.012 | 0.021 | 0.027 | 0.034 | 0.041 | 0.056 | 0.066 | 0.074 | 0.083 | 0.086 | 0.089 | 0.089 | 0.089 | 0.085 |
| Argentina | 0.068 | 0.068 | 0.069 | 0.073 | 0.083 | 0.091 | 0.102 | 0.118 | 0.134 | 0.153 | 0.157 | 0.162 | 0.168 | 0.175 | 0.186 |
| Armenia | 0.081 | 0.097 | 0.114 | 0.148 | 0.183 | 0.207 | 0.223 | 0.259 | 0.288 | 0.301 | 0.324 | 0.332 | 0.331 | 0.339 | 0.343 |
| Austria | 0.429 | 0.399 | 0.403 | 0.403 | 0.407 | 0.408 | 0.407 | 0.438 | 0.450 | 0.453 | 0.523 | 0.515 | 0.488 | 0.487 | 0.492 |
| Azerbaijan | 0.140 | 0.115 | 0.096 | 0.111 | 0.126 | 0.137 | 0.145 | 0.156 | 0.158 | 0.168 | 0.179 | 0.181 | 0.171 | 0.150 | 0.152 |
| Bahamas | 0.333 | 0.289 | 0.307 | 0.361 | 0.358 | 0.349 | 0.345 | 0.366 | 0.363 | 0.429 | 0.442 | 0.427 | 0.438 | 0.414 | 0.418 |
| Bangladesh | 0.163 | 0.165 | 0.170 | 0.176 | 0.183 | 0.196 | 0.215 | 0.241 | 0.254 | 0.274 | 0.293 | 0.318 | 0.337 | 0.349 | 0.364 |
| Belgium | 0.935 | 0.934 | 0.933 | 0.940 | 0.934 | 0.925 | 0.920 | 0.916 | 0.913 | 0.908 | 0.894 | 0.887 | 0.874 | 0.864 | 0.934 |
| Belize | 0.179 | 0.180 | 0.174 | 0.192 | 0.198 | 0.196 | 0.196 | 0.195 | 0.190 | 0.198 | 0.195 | 0.183 | 0.180 | 0.176 | 0.186 |
| Bhutan | 0.071 | 0.068 | 0.065 | 0.067 | 0.069 | 0.071 | 0.087 | 0.108 | 0.077 | 0.119 | 0.122 | 0.131 | 0.143 | 0.161 | 0.119 |
| Bolivia | 0.074 | 0.102 | 0.054 | 0.099 | 0.104 | 0.119 | 0.136 | 0.162 | 0.176 | 0.199 | 0.222 | 0.240 | 0.263 | 0.274 | 0.276 |
| Bosnia and Herzegovina | 0.217 | 0.284 | 0.217 | 0.253 | 0.287 | 0.315 | 0.288 | 0.303 | 0.308 | 0.322 | 0.334 | 0.325 | 0.318 | 0.322 | 0.326 |
| Botswana | 0.060 | 0.061 | 0.068 | 0.098 | 0.096 | 0.093 | 0.097 | 0.091 | 0.091 | 0.094 | 0.098 | 0.101 | 0.102 | 0.102 | 0.123 |
| Brazil | 0.315 | 0.322 | 0.318 | 0.324 | 0.333 | 0.338 | 0.344 | 0.344 | 0.348 | 0.355 | 0.359 | 0.351 | 0.343 | 0.330 | 0.321 |
| Brunei Darussalam | 0.210 | 0.261 | 0.254 | 0.284 | 0.319 | 0.352 | 0.350 | 0.347 | 0.377 | 0.354 | 0.336 | 0.336 | 0.321 | 0.297 | 0.304 |
| Bulgaria | 0.445 | 0.466 | 0.517 | 0.555 | 0.593 | 0.612 | 0.617 | 0.574 | 0.573 | 0.572 | 0.569 | 0.566 | 0.542 | 0.555 | 0.553 |
| Burundi | 0.012 | 0.013 | 0.014 | 0.016 | 0.016 | 0.019 | 0.023 | 0.026 | 0.030 | 0.038 | 0.038 | 0.038 | 0.039 | 0.038 | 0.016 |
| Cambodia | 0.100 | 0.023 | 0.014 | 0.020 | 0.027 | 0.034 | 0.037 | 0.040 | 0.045 | 0.050 | 0.062 | 0.072 | 0.082 | 0.090 | 0.100 |
| Cameroon | 0.002 | 0.005 | 0.006 | 0.006 | 0.008 | 0.009 | 0.011 | 0.013 | 0.014 | 0.016 | 0.017 | 0.018 | 0.019 | 0.019 | 0.021 |
| Central African Republic | 0.000 | 0.000 | 0.000 | 0.001 | 0.001 | 0.003 | 0.004 | 0.005 | 0.005 | 0.006 | 0.004 | 0.003 | 0.004 | 0.004 | 0.004 |
| Chad | 0.000 | 0.001 | 0.000 | 0.001 | 0.002 | 0.002 | 0.002 | 0.003 | 0.003 | 0.004 | 0.005 | 0.006 | 0.006 | 0.006 | 0.006 |
| Chile | 0.144 | 0.159 | 0.175 | 0.190 | 0.204 | 0.208 | 0.219 | 0.225 | 0.231 | 0.223 | 0.206 | 0.201 | 0.194 | 0.186 | 0.178 |
| China | 0.106 | 0.182 | 0.099 | 0.106 | 0.112 | 0.132 | 0.149 | 0.167 | 0.182 | 0.213 | 0.243 | 0.321 | 0.339 | 0.349 | 0.393 |
| Colombia | 0.151 | 0.151 | 0.180 | 0.137 | 0.134 | 0.137 | 0.151 | 0.151 | 0.161 | 0.173 | 0.177 | 0.182 | 0.182 | 0.180 | 0.179 |
| Comoros | 0.007 | 0.006 | 0.006 | 0.014 | 0.014 | 0.018 | 0.020 | 0.053 | 0.050 | 0.054 | 0.055 | 0.054 | 0.059 | 0.060 | 0.062 |
| Congo, Democratic Republic of | 0.002 | 0.002 | 0.002 | 0.002 | 0.002 | 0.001 | 0.003 | 0.003 | 0.004 | 0.003 | 0.005 | 0.007 | 0.002 | 0.002 | 0.004 |
| Congo, Republic of | 0.003 | 0.003 | 0.003 | 0.007 | 0.009 | 0.009 | 0.012 | 0.015 | 0.017 | 0.022 | 0.028 | 0.032 | 0.007 | 0.009 | 0.017 |
| Costa Rica | 0.165 | 0.185 | 0.198 | 0.225 | 0.231 | 0.239 | 0.246 | 0.289 | 0.295 | 0.342 | 0.362 | 0.313 | 0.319 | 0.316 | 0.340 |
| Croatia | 0.382 | 0.370 | 0.394 | 0.433 | 0.469 | 0.497 | 0.508 | 0.520 | 0.527 | 0.524 | 0.526 | 0.530 | 0.528 | 0.532 | 0.522 |
| Czech Republic | 0.261 | 0.274 | 0.288 | 0.288 | 0.299 | 0.305 | 0.310 | 0.322 | 0.333 | 0.347 | 0.348 | 0.348 | 0.344 | 0.345 | 0.350 |
| Djibouti | 0.011 | 0.011 | 0.013 | 0.013 | 0.017 | 0.022 | 0.028 | 0.032 | 0.035 | 0.036 | 0.043 | 0.050 | 0.052 | 0.065 | 0.032 |
| Dominican Republic | 0.144 | 0.162 | 0.165 | 0.172 | 0.177 | 0.183 | 0.188 | 0.199 | 0.203 | 0.222 | 0.226 | 0.246 | 0.254 | 0.260 | 0.251 |
| Ecuador | 0.076 | 0.077 | 0.092 | 0.148 | 0.147 | 0.169 | 0.176 | 0.134 | 0.140 | 0.139 | 0.145 | 0.149 | 0.147 | 0.135 | 0.145 |
| Egypt, Arab Republic of | 0.028 | 0.030 | 0.033 | 0.038 | 0.042 | 0.046 | 0.049 | 0.051 | 0.053 | 0.055 | 0.059 | 0.064 | 0.070 | 0.076 | 0.080 |
| El Salvador | 0.240 | 0.209 | 0.256 | 0.262 | 0.274 | 0.240 | 0.232 | 0.242 | 0.246 | 0.267 | 0.268 | 0.279 | 0.303 | 0.325 | 0.332 |
| Equatorial Guinea | 0.011 | 0.012 | 0.013 | 0.015 | 0.019 | 0.022 | 0.027 | 0.032 | 0.030 | 0.034 | 0.041 | 0.042 | 0.011 | 0.015 |  |
| Estonia | 0.248 | 0.274 | 0.298 | 0.322 | 0.325 | 0.304 | 0.300 | 0.286 | 0.270 | 0.249 | 0.234 | 0.218 | 0.209 | 0.202 | 0.201 |
| Eswatini | 0.051 | 0.063 | 0.070 | 0.072 | 0.082 | 0.082 | 0.091 | 0.098 | 0.107 | 0.110 | 0.126 | 0.132 | 0.129 | 0.123 | 0.134 |
| Fiji | 0.092 | 0.096 | 0.105 | 0.112 | 0.122 | 0.123 | 0.136 | 0.143 | 0.142 | 0.158 | 0.173 | 0.177 | 0.187 | 0.185 | 0.187 |
| Finland | 0.140 | 0.144 | 0.163 | 0.158 | 0.157 | 0.151 | 0.153 | 0.150 | 0.141 | 0.136 | 0.134 | 0.102 | 0.092 | 0.074 | 0.087 |
| Gabon | 0.029 | 0.028 | 0.028 | 0.030 | 0.029 | 0.033 | 0.038 | 0.043 | 0.064 | 0.068 | 0.038 | 0.038 | 0.029 | 0.029 | 0.068 |
| Georgia | 0.064 | 0.082 | 0.115 | 0.164 | 0.230 | 0.226 | 0.240 | 0.260 | 0.315 | 0.354 | 0.358 | 0.354 | 0.367 | 0.373 | 0.367 |
| Ghana | 0.050 | 0.028 | 0.031 | 0.049 | 0.038 | 0.041 | 0.043 | 0.044 | 0.050 | 0.058 | 0.060 | 0.074 | 0.077 | 0.088 | 0.091 |
| Greece | 0.440 | 0.415 | 0.446 | 0.476 | 0.502 | 0.499 | 0.488 | 0.467 | 0.449 | 0.384 | 0.347 | 0.336 | 0.323 | 0.318 | 0.307 |
| Guatemala | 0.192 | 0.177 | 0.175 | 0.251 | 0.258 | 0.282 | 0.296 | 0.310 | 0.318 | 0.340 | 0.347 | 0.352 | 0.352 | 0.350 | 0.304 |
| Guinea | 0.003 | 0.003 | 0.003 | 0.004 | 0.005 | 0.007 | 0.008 | 0.009 | 0.010 | 0.011 | 0.015 | 0.018 | 0.018 | 0.018 | 0.019 |
| Guyana | 0.047 | 0.049 | 0.051 | 0.054 | 0.056 | 0.057 | 0.059 | 0.063 | 0.066 | 0.068 | 0.070 | 0.072 | 0.076 | 0.078 | 0.078 |
| Haiti | 0.041 | 0.042 | 0.041 | 0.041 | 0.036 | 0.036 | 0.036 | 0.038 | 0.040 | 0.035 | 0.036 | 0.034 | 0.036 | 0.039 | 0.041 |
| Honduras | 0.095 | 0.103 | 0.118 | 0.145 | 0.160 | 0.162 | 0.149 | 0.164 | 0.177 | 0.183 | 0.183 | 0.174 | 0.178 | 0.178 | 0.177 |
| Hungary | 0.215 | 0.230 | 0.250 | 0.275 | 0.292 | 0.293 | 0.290 | 0.290 | 0.283 | 0.280 | 0.283 | 0.278 | 0.282 | 0.285 | 0.281 |
| Iceland | 0.448 | 0.458 | 0.448 | 0.448 | 0.439 | 0.395 | 0.399 | 0.388 | 0.364 | 0.349 | 0.341 | 0.313 | 0.338 | 0.334 | 0.315 |
| India | 0.227 | 0.113 | 0.116 | 0.122 | 0.132 | 0.142 | 0.160 | 0.176 | 0.198 | 0.219 | 0.262 | 0.285 | 0.307 | 0.319 | 0.320 |
| Indonesia | 0.060 | 0.063 | 0.070 | 0.075 | 0.085 | 0.096 | 0.096 | 0.155 | 0.231 | 0.258 | 0.283 | 0.296 | 0.299 | 0.300 | 0.295 |
| Ireland | 0.431 | 0.422 | 0.422 | 0.428 | 0.440 | 0.442 | 0.396 | 0.384 | 0.360 | 0.350 | 0.380 | 0.364 | 0.360 | 0.354 | 0.327 |
| Italy | 0.778 | 0.787 | 0.810 | 0.840 | 0.888 | 0.919 | 0.893 | 0.899 | 0.891 | 0.856 | 0.845 | 0.852 | 0.822 | 0.787 | 0.747 |
| Jamaica | 0.133 | 0.138 | 0.148 | 0.149 | 0.157 | 0.161 | 0.162 | 0.167 | 0.169 | 0.170 | 0.176 | 0.181 | 0.187 | 0.210 | 0.220 |
| Japan | 0.922 | 0.921 | 0.920 | 0.919 | 0.918 | 0.918 | 0.918 | 0.919 | 0.919 | 0.919 | 0.919 | 0.920 | 0.920 | 0.919 | 0.919 |
| Jordan | 0.168 | 0.164 | 0.158 | 0.148 | 0.153 | 0.154 | 0.161 | 0.163 | 0.163 | 0.163 | 0.164 | 0.162 | 0.170 | 0.171 | 0.154 |
| Kenya | 0.017 | 0.017 | 0.021 | 0.029 | 0.036 | 0.041 | 0.046 | 0.049 | 0.052 | 0.053 | 0.055 | 0.056 | 0.054 | 0.054 | 0.053 |
| Kiribati | 0.080 | 0.069 | 0.069 | 0.069 | 0.080 | 0.080 | 0.080 | 0.070 | 0.069 | 0.080 | 0.069 | 0.069 | 0.069 | 0.069 | 0.069 |
| Korea, Republic of | 0.758 | 0.767 | 0.777 | 0.787 | 0.796 | 0.790 | 0.792 | 0.796 | 0.799 | 0.796 | 0.785 | 0.780 | 0.773 | 0.760 | 0.760 |
| Kosobo | 0.322 | 0.214 | 0.216 | 0.225 | 0.246 | 0.270 | 0.287 | 0.297 | 0.301 | 0.295 | 0.284 | 0.295 | 0.287 | 0.259 | 0.255 |
| Lao People's Democratic Republic | 0.056 | 0.019 | 0.056 | 0.030 | 0.019 | 0.021 | 0.030 | 0.035 | 0.040 | 0.051 | 0.056 | 0.063 | 0.065 | 0.070 | 0.070 |
| Latvia | 0.275 | 0.277 | 0.294 | 0.340 | 0.352 | 0.352 | 0.353 | 0.329 | 0.295 | 0.268 | 0.248 | 0.236 | 0.226 | 0.224 | 0.212 |

Table A.2. *Financial outreach index, cont.*

| Country | 2004 | 2005 | 2006 | 2007 | 2008 | 2009 | 2010 | 2011 | 2012 | 2013 | 2014 | 2015 | 2016 | 2017 | 2018 |
|---|---|---|---|---|---|---|---|---|---|---|---|---|---|---|---|
| Lebanon | 0.494 | 0.509 | 0.518 | 0.534 | 0.548 | 0.565 | 0.581 | 0.593 | 0.601 | 0.606 | 0.613 | 0.624 | 0.633 | 0.646 | 0.660 |
| Lesotho | 0.019 | 0.021 | 0.019 | 0.022 | 0.025 | 0.028 | 0.036 | 0.038 | 0.041 | 0.047 | 0.048 | 0.054 | 0.059 | 0.058 | 0.058 |
| Liberia | 0.021 | 0.005 | 0.005 | 0.007 | 0.010 | 0.017 | 0.021 | 0.021 | 0.021 | 0.022 | 0.023 | 0.023 | 0.021 | 0.017 | 0.023 |
| Libyan Arab Jamahirya | 0.047 | 0.047 | 0.055 | 0.056 | 0.055 | 0.056 | 0.058 | 0.058 | 0.060 | 0.063 | 0.063 | 0.062 | 0.062 | 0.061 | 0.055 |
| Madagascar | 0.005 | 0.006 | 0.006 | 0.007 | 0.008 | 0.009 | 0.009 | 0.010 | 0.011 | 0.011 | 0.013 | 0.015 | 0.016 | 0.017 | 0.006 |
| Malawi | 0.007 | 0.009 | 0.009 | 0.016 | 0.018 | 0.021 | 0.027 | 0.017 | 0.034 | 0.035 | 0.036 | 0.038 | 0.009 | 0.027 | 0.017 |
| Malaysia | 0.152 | 0.145 | 0.143 | 0.169 | 0.176 | 0.204 | 0.208 | 0.212 | 0.212 | 0.216 | 0.210 | 0.207 | 0.200 | 0.196 | 0.195 |
| Maldives | 0.304 | 0.317 | 0.369 | 0.506 | 0.510 | 0.521 | 0.522 | 0.529 | 0.576 | 0.599 | 0.630 | 0.629 | 0.644 | 0.658 | 0.663 |
| Malta | 0.804 | 0.805 | 0.789 | 0.804 | 0.813 | 0.816 | 0.819 | 0.834 | 0.827 | 0.824 | 0.814 | 0.815 | 0.801 | 0.790 | 0.775 |
| Mauritania | 0.025 | 0.057 | 0.057 | 0.025 | 0.036 | 0.037 | 0.037 | 0.026 | 0.029 | 0.041 | 0.045 | 0.050 | 0.057 | 0.057 | 0.037 |
| Mauritius | 0.524 | 0.551 | 0.569 | 0.609 | 0.619 | 0.654 | 0.668 | 0.691 | 0.702 | 0.709 | 0.714 | 0.717 | 0.704 | 0.680 | 0.655 |
| Mexico | 0.129 | 0.136 | 0.146 | 0.163 | 0.175 | 0.182 | 0.191 | 0.191 | 0.206 | 0.206 | 0.211 | 0.213 | 0.218 | 0.221 | 0.229 |
| Micronesia, Federated States of | 0.095 | 0.095 | 0.095 | 0.096 | 0.096 | 0.118 | 0.136 | 0.134 | 0.151 | 0.149 | 0.147 | 0.145 | 0.143 | 0.141 | 0.140 |
| Moldova | 0.230 | 0.256 | 0.270 | 0.297 | 0.337 | 0.343 | 0.354 | 0.382 | 0.392 | 0.405 | 0.425 | 0.290 | 0.295 | 0.298 | 0.302 |
| Mongolia | 0.325 | 0.212 | 0.274 | 0.408 | 0.258 | 0.268 | 0.275 | 0.300 | 0.325 | 0.335 | 0.352 | 0.379 | 0.409 | 0.418 | 0.444 |
| Montenegro | 0.300 | 0.173 | 0.212 | 0.287 | 0.358 | 0.384 | 0.373 | 0.381 | 0.391 | 0.406 | 0.402 | 0.420 | 0.432 | 0.435 | 0.449 |
| Morocco | 0.077 | 0.098 | 0.098 | 0.109 | 0.124 | 0.162 | 0.173 | 0.184 | 0.195 | 0.204 | 0.210 | 0.216 | 0.220 | 0.222 | 0.225 |
| Mozambique | 0.012 | 0.014 | 0.016 | 0.019 | 0.021 | 0.023 | 0.028 | 0.031 | 0.033 | 0.035 | 0.039 | 0.043 | 0.044 | 0.044 | 0.044 |
| Myanmar | 0.016 | 0.024 | 0.012 | 0.014 | 0.013 | 0.015 | 0.008 | 0.009 | 0.010 | 0.017 | 0.024 | 0.024 | 0.027 | 0.039 | 0.045 |
| Namibia | 0.067 | 0.129 | 0.072 | 0.076 | 0.114 | 0.138 | 0.153 | 0.156 | 0.151 | 0.158 | 0.165 | 0.167 | 0.192 | 0.165 | 0.067 |
| Nepal | 0.042 | 0.056 | 0.049 | 0.064 | 0.050 | 0.072 | 0.071 | 0.081 | 0.091 | 0.095 | 0.100 | 0.106 | 0.116 | 0.135 | 0.176 |
| Netherlands | 0.804 | 0.756 | 0.780 | 0.791 | 0.786 | 0.772 | 0.741 | 0.710 | 0.674 | 0.630 | 0.596 | 0.559 | 0.523 | 0.498 | 0.461 |
| Nicaragua | 0.032 | 0.042 | 0.049 | 0.054 | 0.060 | 0.061 | 0.054 | 0.059 | 0.063 | 0.072 | 0.077 | 0.085 | 0.101 | 0.101 | 0.102 |
| North Macedonia | 0.271 | 0.129 | 0.166 | 0.212 | 0.277 | 0.294 | 0.302 | 0.293 | 0.292 | 0.305 | 0.310 | 0.323 | 0.323 | 0.326 | 0.318 |
| Norway | 0.149 | 0.177 | 0.179 | 0.181 | 0.177 | 0.170 | 0.165 | 0.163 | 0.155 | 0.146 | 0.139 | 0.131 | 0.109 | 0.101 | 0.108 |
| Oman | 0.176 | 0.165 | 0.172 | 0.172 | 0.177 | 0.177 | 0.178 | 0.172 | 0.166 | 0.167 | 0.154 | 0.149 | 0.147 | 0.147 | 0.142 |
| Pakistan | 0.060 | 0.062 | 0.067 | 0.072 | 0.078 | 0.082 | 0.084 | 0.088 | 0.094 | 0.101 | 0.109 | 0.118 | 0.126 | 0.132 | 0.136 |
| Panama | 0.211 | 0.292 | 0.203 | 0.192 | 0.203 | 0.211 | 0.217 | 0.230 | 0.242 | 0.262 | 0.278 | 0.292 | 0.296 | 0.293 | 0.290 |
| Papua New Guinea | 0.015 | 0.015 | 0.014 | 0.015 | 0.016 | 0.016 | 0.017 | 0.021 | 0.020 | 0.021 | 0.022 | 0.021 | 0.025 | 0.024 | 0.023 |
| Paraguay | 0.070 | 0.070 | 0.056 | 0.071 | 0.078 | 0.059 | 0.079 | 0.084 | 0.089 | 0.097 | 0.092 | 0.100 | 0.099 | 0.104 | 0.109 |
| Peru | 0.042 | 0.044 | 0.050 | 0.059 | 0.074 | 0.082 | 0.090 | 0.101 | 0.120 | 0.127 | 0.160 | 0.291 | 0.276 | 0.269 | 0.280 |
| Philippines | 0.117 | 0.121 | 0.124 | 0.124 | 0.129 | 0.134 | 0.141 | 0.152 | 0.166 | 0.185 | 0.197 | 0.210 | 0.222 | 0.232 | 0.240 |
| Poland | 0.275 | 0.282 | 0.301 | 0.332 | 0.378 | 0.397 | 0.398 | 0.411 | 0.429 | 0.428 | 0.444 | 0.446 | 0.458 | 0.443 | 0.443 |
| Portugal | 0.796 | 0.815 | 0.833 | 0.825 | 0.845 | 0.855 | 0.860 | 0.851 | 0.836 | 0.821 | 0.804 | 0.679 | 0.725 | 0.698 | 0.674 |
| Romania | 0.253 | 0.242 | 0.296 | 0.284 | 0.392 | 0.396 | 0.398 | 0.411 | 0.398 | 0.384 | 0.381 | 0.377 | 0.367 | 0.357 | 0.343 |
| Rwanda | 0.003 | 0.008 | 0.007 | 0.011 | 0.050 | 0.056 | 0.060 | 0.074 | 0.087 | 0.095 | 0.097 | 0.102 | 0.105 | 0.105 | 0.099 |
| Samoa | 0.127 | 0.131 | 0.145 | 0.172 | 0.175 | 0.200 | 0.203 | 0.168 | 0.174 | 0.214 | 0.238 | 0.222 | 0.245 | 0.251 | 0.260 |
| San Marino | 1.000 | 1.000 | 1.000 | 1.000 | 1.000 | 1.000 | 1.000 | 1.000 | 1.000 | 1.000 | 1.000 | 1.000 | 1.000 | 1.000 | 1.000 |
| Sao Tome and Principe | | | | | | | | 0.275 | 0.309 | 0.312 | 0.313 | 0.317 | 0.273 | 0.284 | 0.253 |
| Saudi Arabia | 0.092 | 0.095 | 0.111 | 0.126 | 0.138 | 0.147 | 0.153 | 0.157 | 0.161 | 0.168 | 0.179 | 0.190 | 0.192 | 0.192 | 0.191 |
| Serbia | 0.271 | 0.244 | 0.295 | 0.356 | 0.405 | 0.406 | 0.394 | 0.384 | 0.367 | 0.338 | 0.316 | 0.314 | 0.326 | 0.303 | 0.308 |
| Seychelles | 0.464 | 0.448 | 0.466 | 0.469 | 0.548 | 0.555 | 0.563 | 0.569 | 0.599 | 0.657 | 0.703 | 0.711 | 0.736 | 0.765 | 0.770 |
| Solomon Islands | 0.038 | 0.041 | 0.047 | 0.049 | 0.050 | 0.050 | 0.044 | 0.042 | 0.042 | 0.039 | 0.042 | 0.042 | 0.043 | 0.045 | 0.042 |
| South Africa | 0.090 | 0.092 | 0.097 | 0.099 | 0.138 | 0.164 | 0.178 | 0.185 | 0.183 | 0.187 | 0.208 | 0.214 | 0.213 | 0.212 | 0.209 |
| South Sudan | | | | | 0.006 | 0.003 | 0.005 | 0.004 | 0.005 | 0.007 | 0.010 | 0.009 | 0.008 | 0.007 | 0.006 |
| Spain | 0.780 | 0.792 | 0.806 | 0.821 | 0.825 | 0.814 | 0.805 | 0.788 | 0.768 | 0.740 | 0.720 | 0.711 | 0.689 | 0.679 | 0.664 |
| Suriname | 0.082 | 0.087 | 0.096 | 0.104 | 0.102 | 0.114 | 0.119 | 0.121 | 0.129 | 0.137 | 0.146 | 0.140 | 0.146 | 0.150 | 0.145 |
| Sweden | 0.197 | 0.199 | 0.200 | 0.206 | 0.211 | 0.208 | 0.206 | 0.207 | 0.203 | 0.196 | 0.194 | 0.185 | 0.164 | 0.151 | 0.167 |
| Switzerland | 0.803 | 0.811 | 0.818 | 0.829 | 0.833 | 0.837 | 0.842 | 0.847 | 0.841 | 0.831 | 0.818 | 0.811 | 0.797 | 0.783 | 0.778 |
| Syrian Arab Republic | 0.014 | 0.015 | 0.016 | 0.017 | 0.025 | 0.041 | 0.045 | 0.048 | 0.050 | 0.028 | 0.050 | 0.045 | 0.025 | 0.017 | 0.050 |
| Tanzania | 0.013 | 0.016 | 0.019 | 0.013 | 0.013 | 0.016 | 0.018 | 0.020 | 0.022 | 0.025 | 0.025 | 0.026 | 0.026 | 0.013 | 0.013 |
| Thailand | 0.119 | 0.154 | 0.197 | 0.224 | 0.280 | 0.310 | 0.340 | 0.360 | 0.388 | 0.416 | 0.446 | 0.455 | 0.456 | 0.467 | 0.461 |
| Tonga | 0.232 | 0.209 | 0.199 | 0.208 | 0.250 | 0.237 | 0.232 | 0.233 | 0.223 | 0.235 | 0.209 | 0.297 | 0.318 | 0.363 | 0.354 |
| Trinidad and Tobago | | | | | 0.263 | 0.274 | 0.285 | 0.282 | 0.291 | 0.301 | 0.311 | 0.310 | 0.315 | 0.313 | 0.312 |
| Turkey | 0.184 | 0.161 | 0.177 | 0.197 | 0.224 | 0.235 | 0.256 | 0.278 | 0.294 | 0.326 | 0.341 | 0.348 | 0.339 | 0.336 | 0.336 |
| Uganda | 0.009 | 0.011 | 0.012 | 0.014 | 0.020 | 0.025 | 0.026 | 0.027 | 0.030 | 0.034 | 0.034 | 0.035 | 0.034 | 0.032 | 0.032 |
| Ukraine | 0.078 | 0.100 | 0.124 | 0.166 | 0.214 | 0.218 | 0.221 | 0.237 | 0.256 | 0.285 | 0.259 | 0.236 | 0.240 | 0.263 | 0.262 |
| United Arab Emirates | 0.132 | 0.186 | 0.179 | 0.188 | 0.202 | 0.244 | 0.245 | 0.259 | 0.273 | 0.282 | 0.292 | 0.303 | 0.304 | 0.297 | 0.291 |
| United Kingdom | 0.792 | 0.786 | 0.769 | 0.770 | 0.767 | 0.762 | 0.756 | 0.750 | 0.731 | 0.764 | 0.792 | 0.792 | 0.792 | 0.792 | 0.746 |
| Uzbekistan | 0.224 | 0.231 | 0.245 | 0.241 | 0.240 | 0.239 | 0.240 | 0.282 | 0.285 | 0.299 | 0.279 | 0.212 | 0.223 | 0.293 | 0.291 |
| Vanuatu | 0.105 | 0.105 | 0.105 | 0.122 | 0.139 | 0.145 | 0.158 | 0.160 | 0.178 | 0.180 | 0.180 | 0.184 | 0.182 | 0.196 | 0.194 |
| Venezuela, Republica Bolivariana | 0.171 | 0.128 | 0.136 | 0.144 | 0.147 | 0.154 | 0.171 | 0.176 | 0.178 | 0.183 | 0.183 | 0.186 | 0.147 | 0.186 | 0.178 |
| Vietnam | 0.038 | 0.054 | 0.055 | 0.073 | 0.084 | 0.098 | 0.108 | 0.124 | 0.125 | 0.137 | 0.144 | 0.149 | 0.153 | 0.151 | 0.160 |
| West Bank and Gaza | 0.177 | 0.281 | 0.160 | 0.182 | 0.210 | 0.238 | 0.244 | 0.263 | 0.279 | 0.294 | 0.321 | 0.340 | 0.366 | 0.387 | 0.475 |
| Yemen, Republic of | 0.009 | 0.010 | 0.012 | 0.013 | 0.014 | 0.015 | 0.016 | 0.016 | 0.017 | 0.018 | 0.022 | 0.022 | 0.022 | 0.022 | 0.022 |
| Zambia | 0.015 | 0.015 | 0.017 | 0.020 | 0.025 | 0.030 | 0.031 | 0.033 | 0.037 | 0.040 | 0.043 | 0.045 | 0.045 | 0.045 | 0.042 |
| Zimbabwe | 0.024 | 0.034 | 0.033 | 0.036 | 0.032 | 0.039 | 0.035 | 0.040 | 0.076 | 0.079 | 0.087 | 0.044 | 0.043 | 0.042 | 0.041 |

Source: Own estimates using Financial Access Survey, 2004-2018 (IMF, 2019). A few countries are missing in these tables as some of their information was not available at the time of writing this paper.

Table A.3. *Financial usage index*

| Country | 2004 | 2005 | 2006 | 2007 | 2008 | 2009 | 2010 | 2011 | 2012 | 2013 | 2014 | 2015 | 2016 | 2017 | 2018 |
|---|---|---|---|---|---|---|---|---|---|---|---|---|---|---|---|
| Afghanistan | 0.043 | 0.044 | 0.028 | 0.032 | 0.044 | 0.028 | 0.028 | 0.025 | 0.040 | 0.032 | 0.043 | 0.045 | 0.046 | 0.043 | 0.044 |
| Albania | 0.477 | 0.409 | 0.389 | 0.477 | 0.409 | 0.389 | 0.477 | 0.389 | 0.477 | 0.477 | 0.477 | 0.409 | 0.477 | 0.409 | 0.389 |
| Algeria | 0.159 | 0.161 | 0.169 | 0.181 | 0.181 | 0.116 | 0.133 | 0.138 | 0.147 | 0.151 | 0.152 | 0.150 | 0.146 | 0.147 | 0.147 |
| Angola | 0.006 | 0.004 | 0.004 | 0.004 | 0.004 | 0.004 | 0.004 | 0.004 | 0.004 | 0.004 | 0.004 | 0.004 | 0.004 | 0.006 | 0.005 |
| Argentina | 0.202 | 0.232 | 0.248 | 0.276 | 0.295 | 0.312 | 0.329 | 0.352 | 0.395 | 0.404 | 0.426 | 0.483 | 0.534 | 0.580 | 0.618 |
| Armenia | 0.114 | 0.179 | 0.182 | 0.177 | 0.180 | 0.161 | 0.191 | 0.234 | 0.282 | 0.336 | 0.410 | 0.438 | 0.474 | 0.480 | 0.537 |
| Austria | 0.340 | 0.348 | 0.345 | 0.350 | 0.350 | 0.341 | 0.334 | 0.327 | 0.313 | 0.319 | 0.312 | 0.305 | 0.299 | 0.272 | 0.283 |
| Azerbaijan | 0.577 | 0.080 | 0.201 | 0.080 | 0.080 | 0.080 | 0.111 | 0.140 | 0.201 | 0.303 | 0.465 | 0.535 | 0.495 | 0.522 | 0.577 |
| Bahamas | 0.638 | 0.663 | 0.688 | 0.720 | 0.723 | 0.696 | 0.628 | 0.617 | 0.582 | 0.547 | 0.522 | 0.496 | 0.522 | 0.517 | 0.513 |
| Bangladesh | 0.108 | 0.111 | 0.114 | 0.115 | 0.120 | 0.121 | 0.141 | 0.156 | 0.164 | 0.175 | 0.180 | 0.193 | 0.205 | 0.217 | 0.233 |
| Belgium | 1.000 | 1.000 | 1.000 | 1.000 | 1.000 | 1.000 | 1.000 | 1.000 | 1.000 | 1.000 | 1.000 | 1.000 | 1.000 | 1.000 | 1.000 |
| Belize | 0.227 | 0.248 | 0.299 | 0.248 | 0.265 | 0.299 | 0.227 | 0.299 | 0.286 | 0.280 | 0.265 | 0.241 | 0.248 | 0.227 | 0.234 |
| Bhutan | 0.121 | 0.109 | 0.115 | 0.121 | 0.133 | 0.147 | 0.197 | 0.259 | 0.118 | 0.270 | 0.361 | 0.397 | 0.443 | 0.382 | 0.270 |
| Bolivia | 0.036 | 0.044 | 0.035 | 0.040 | 0.049 | 0.058 | 0.069 | 0.081 | 0.093 | 0.108 | 0.124 | 0.141 | 0.158 | 0.191 | 0.204 |
| Bosnia and Herzegovina | 0.270 | 0.267 | 0.332 | 0.329 | 0.348 | 0.356 | 0.353 | 0.376 | 0.389 | 0.417 | 0.415 | 0.420 | 0.416 | 0.418 | 0.428 |
| Botswana | 0.290 | 0.290 | 0.319 | 0.292 | 0.290 | 0.319 | 0.319 | 0.283 | 0.290 | 0.292 | 0.291 | 0.266 | 0.264 | 0.268 | 0.319 |
| Brazil | 0.836 | 0.886 | 0.853 | 0.277 | 0.311 | 0.351 | 0.395 | 0.450 | 0.766 | 0.838 | 0.831 | 0.804 | 0.909 | 0.932 | 0.887 |
| Brunei Darussalam | 0.667 | 0.603 | 0.680 | 0.673 | 0.680 | 0.699 | 0.673 | 0.667 | 0.818 | 0.773 | 0.680 | 0.682 | 0.603 | 0.597 | 0.594 |
| Bulgaria | 0.604 | 0.436 | 0.487 | 0.549 | 0.609 | 0.616 | 0.597 | 0.604 | 0.609 | 0.595 | 0.583 | 0.561 | 0.551 | 0.540 | 0.547 |
| Burundi | 0.004 | 0.005 | 0.005 | 0.005 | 0.006 | 0.007 | 0.009 | 0.010 | 0.013 | 0.011 | 0.012 | 0.013 | 0.011 | 0.011 | 0.005 |
| Cambodia | 0.108 | 0.031 | 0.084 | 0.108 | 0.025 | 0.031 | 0.035 | 0.039 | 0.045 | 0.051 | 0.063 | 0.073 | 0.084 | 0.093 | 0.108 |
| Cameroon | 0.000 | 0.000 | 0.006 | 0.013 | 0.017 | 0.020 | 0.022 | 0.025 | 0.028 | 0.030 | 0.026 | 0.032 | 0.033 | 0.033 | 0.049 |
| Central African Republic | 0.007 | 0.001 | 0.001 | 0.004 | 0.007 | 0.009 | 0.010 | 0.012 | 0.013 | 0.019 | 0.015 | 0.017 | 0.020 | 0.019 | 0.015 |
| Chad | 0.011 | 0.011 | 0.012 | 0.010 | 0.011 | 0.011 | 0.011 | 0.011 | 0.011 | 0.009 | 0.010 | 0.011 | 0.012 | 0.011 | 0.011 |
| Chile | 0.483 | 0.505 | 0.575 | 0.628 | 0.674 | 0.704 | 0.744 | 0.800 | 0.837 | 0.870 | 0.882 | 0.891 | 0.905 | 0.925 | 0.948 |
| China | 0.003 | 0.005 | 0.002 | 0.003 | 0.003 | 0.003 | 0.004 | 0.005 | 0.005 | 0.006 | 0.006 | 0.007 | 0.008 | 0.009 | 0.010 |
| Colombia | 0.291 | 0.324 | 0.358 | 0.369 | 0.376 | 0.408 | 0.383 | 0.417 | 0.442 | 0.463 | 0.496 | 0.526 | 0.535 | 0.562 | 0.588 |
| Comoros | 0.036 | 0.017 | 0.027 | 0.036 | 0.027 | 0.032 | 0.017 | 0.024 | 0.027 | 0.030 | 0.032 | 0.034 | 0.036 | 0.036 | 0.037 |
| Congo, Democratic Republic of | 0.009 | 0.005 | 0.005 | 0.004 | 0.004 | 0.004 | 0.004 | 0.005 | 0.009 | 0.014 | 0.014 | 0.004 | 0.005 | 0.004 | 0.009 |
| Congo, Republic of | 0.003 | 0.005 | 0.005 | 0.011 | 0.012 | 0.015 | 0.017 | 0.019 | 0.023 | 0.024 | 0.034 | 0.039 | 0.011 | 0.012 | 0.023 |
| Costa Rica | 0.485 | 0.442 | 0.417 | 0.456 | 0.470 | 0.463 | 0.458 | 0.473 | 0.485 | 0.514 | 0.534 | 0.570 | 0.676 | 0.745 | 0.752 |
| Croatia | 0.164 | 0.160 | 0.164 | 0.164 | 0.223 | 0.164 | 0.314 | 0.301 | 0.271 | 0.271 | 0.232 | 0.223 | 0.203 | 0.160 | 0.164 |
| Czech Republic | 0.330 | 0.341 | 0.345 | 0.385 | 0.410 | 0.426 | 0.444 | 0.472 | 0.495 | 0.534 | 0.577 | 0.393 | 0.400 | 0.411 | 0.424 |
| Djibouti | 0.025 | 0.072 | 0.025 | 0.051 | 0.016 | 0.021 | 0.025 | 0.028 | 0.031 | 0.049 | 0.056 | 0.065 | 0.051 | 0.072 | 0.028 |
| Dominican Republic | 0.288 | 0.341 | 0.228 | 0.244 | 0.274 | 0.238 | 0.253 | 0.284 | 0.251 | 0.273 | 0.288 | 0.321 | 0.336 | 0.329 | 0.341 |
| Ecuador | 0.229 | 0.171 | 0.136 | 0.146 | 0.180 | 0.186 | 0.202 | 0.234 | 0.222 | 0.211 | 0.236 | 0.240 | 0.242 | 0.257 | 0.257 |
| Egypt, Arab Republic of | 0.135 | 0.121 | 0.121 | 0.121 | 0.124 | 0.126 | 0.110 | 0.109 | 0.113 | 0.135 | 0.138 | 0.140 | 0.148 | 0.194 | 0.211 |
| El Salvador | 0.253 | 0.251 | 0.267 | 0.237 | 0.246 | 0.281 | 0.280 | 0.286 | 0.287 | 0.292 | 0.293 | 0.294 | 0.294 | 0.329 | 0.331 |
| Equatorial Guinea | 0.022 | 0.025 | 0.028 | 0.033 | 0.034 | 0.037 | 0.038 | 0.043 | 0.047 | 0.045 | 0.050 | 0.054 | 0.056 | 0.022 | 0.033 |
| Estonia | 0.899 | 0.776 | 0.887 | 1.000 | 0.899 | 0.900 | 0.949 | 0.927 | 0.747 | 0.755 | 0.760 | 0.729 | 0.727 | 0.729 | 0.759 |
| Eswatini | 0.201 | 0.204 | 0.201 | 0.165 | 0.202 | 0.201 | 0.204 | 0.165 | 0.192 | 0.202 | 0.221 | 0.223 | 0.201 | 0.204 | 0.201 |
| Fiji | 0.180 | 0.223 | 0.288 | 0.244 | 0.265 | 0.264 | 0.279 | 0.295 | 0.292 | 0.314 | 0.355 | 0.394 | 0.439 | 0.477 | 0.466 |
| Finland | 0.550 | 0.458 | 0.271 | 0.562 | 0.271 | 0.271 | 0.565 | 0.562 | 0.565 | 0.550 | 0.544 | 0.573 | 0.541 | 0.271 | 0.458 |
| Gabon | 0.055 | 0.025 | 0.027 | 0.031 | 0.055 | 0.055 | 0.059 | 0.067 | 0.084 | 0.155 | 0.059 | 0.059 | 0.055 | 0.055 | 0.155 |
| Georgia | 0.083 | 0.127 | 0.175 | 0.227 | 0.290 | 0.297 | 0.333 | 0.407 | 0.470 | 0.516 | 0.624 | 0.638 | 0.708 | 0.813 | 0.838 |
| Ghana | 0.200 | 0.044 | 0.051 | 0.054 | 0.069 | 0.071 | 0.093 | 0.106 | 0.116 | 0.120 | 0.121 | 0.165 | 0.155 | 0.186 | 0.200 |
| Greece | 1.000 | 1.000 | 1.000 | 1.000 | 1.000 | 1.000 | 1.000 | 1.000 | 1.000 | 1.000 | 1.000 | 1.000 | 1.000 | 1.000 | 1.000 |
| Guatemala | 0.224 | 0.252 | 0.258 | 0.288 | 0.325 | 0.337 | 0.357 | 0.396 | 0.430 | 0.462 | 0.485 | 0.487 | 0.467 | 0.438 | 0.395 |
| Guinea | 0.005 | 0.006 | 0.008 | 0.009 | 0.011 | 0.013 | 0.015 | 0.017 | 0.020 | 0.022 | 0.024 | 0.025 | 0.023 | 0.034 | 0.035 |
| Guyana | 0.209 | 0.219 | 0.172 | 0.210 | 0.209 | 0.254 | 0.260 | 0.268 | 0.273 | 0.278 | 0.274 | 0.267 | 0.258 | 0.252 | 0.245 |
| Haiti | 0.052 | 0.059 | 0.068 | 0.075 | 0.082 | 0.088 | 0.084 | 0.084 | 0.083 | 0.077 | 0.072 | 0.072 | 0.069 | 0.071 | 0.076 |
| Honduras | 0.215 | 0.209 | 0.220 | 0.238 | 0.235 | 0.241 | 0.232 | 0.237 | 0.259 | 0.262 | 0.287 | 0.310 | 0.317 | 0.313 | 0.318 |
| Hungary | 0.383 | 0.377 | 0.358 | 0.376 | 0.439 | 0.446 | 0.452 | 0.454 | 0.450 | 0.451 | 0.427 | 0.413 | 0.414 | 0.420 | 0.417 |
| Iceland | 1.000 | 1.000 | 1.000 | 1.000 | 1.000 | 1.000 | 1.000 | 1.000 | 1.000 | 1.000 | 1.000 | 1.000 | 1.000 | 1.000 | 1.000 |
| India | 0.170 | 0.172 | 0.177 | 0.188 | 0.205 | 0.226 | 0.245 | 0.262 | 0.286 | 0.319 | 0.364 | 0.415 | 0.466 | 0.507 | 0.526 |
| Indonesia | 0.149 | 0.157 | 0.147 | 0.150 | 0.159 | 0.168 | 0.190 | 0.206 | 0.225 | 0.263 | 0.274 | 0.283 | 0.312 | 0.437 | 0.444 |
| Ireland | 0.426 | 0.423 | 0.426 | 0.536 | 0.541 | 0.526 | 0.547 | 0.503 | 0.491 | 0.472 | 0.598 | 0.627 | 0.630 | 0.567 | 0.553 |
| Italy | 0.303 | 0.305 | 0.309 | 0.313 | 0.340 | 0.303 | 0.318 | 0.326 | 0.321 | 0.320 | 0.324 | 0.352 | 0.360 | 0.372 | 0.363 |
| Jamaica | 0.362 | 0.348 | 0.335 | 0.320 | 0.315 | 0.308 | 0.298 | 0.281 | 0.293 | 0.297 | 0.308 | 0.312 | 0.318 | 0.394 | 0.411 |
| Japan | 1.000 | 1.000 | 1.000 | 1.000 | 1.000 | 1.000 | 1.000 | 1.000 | 1.000 | 1.000 | 1.000 | 1.000 | 1.000 | 1.000 | 1.000 |
| Jordan | 0.224 | 0.231 | 0.238 | 0.238 | 0.235 | 0.240 | 0.191 | 0.188 | 0.186 | 0.175 | 0.167 | 0.166 | 0.175 | 0.180 | 0.240 |
| Kenya | 0.078 | 0.048 | 0.058 | 0.100 | 0.098 | 0.109 | 0.150 | 0.161 | 0.172 | 0.231 | 0.295 | 0.361 | 0.415 | 0.447 | 0.497 |
| Kiribati | 0.064 | 0.055 | 0.064 | 0.064 | 0.064 | 0.064 | 0.064 | 0.050 | 0.055 | 0.064 | 0.055 | 0.055 | 0.055 | 0.055 | 0.055 |
| Korea, Republic of | 1.000 | 0.976 | 1.000 | 1.000 | 1.000 | 1.000 | 1.000 | 1.000 | 1.000 | 1.000 | 1.000 | 1.000 | 1.000 | 1.000 | 1.000 |
| Kosobo | 0.149 | 0.180 | 0.207 | 0.244 | 0.284 | 0.292 | 0.295 | 0.364 | 0.379 | 0.370 | 0.368 | 0.349 | 0.377 | 0.327 | 0.292 |
| Lao People's Democratic Republic | 0.124 | 0.124 | 0.124 | 0.147 | 0.124 | 0.139 | 0.147 | 0.172 | 0.124 | 0.102 | 0.124 | 0.139 | 0.147 | 0.161 | 0.172 |
| Latvia | 0.405 | 0.469 | 0.534 | 0.707 | 0.759 | 0.743 | 0.747 | 0.801 | 0.784 | 0.783 | 0.661 | 0.628 | 0.606 | 0.641 | 0.634 |

Table A.3. *Financial usage index, cont.*

| Country | 2004 | 2005 | 2006 | 2007 | 2008 | 2009 | 2010 | 2011 | 2012 | 2013 | 2014 | 2015 | 2016 | 2017 | 2018 |
|---|---|---|---|---|---|---|---|---|---|---|---|---|---|---|---|
| Lebanon | 0.411 | 0.280 | 0.296 | 0.318 | 0.360 | 0.391 | 0.411 | 0.403 | 0.411 | 0.372 | 0.341 | 0.330 | 0.320 | 0.325 | 0.332 |
| Lesotho | 0.118 | 0.086 | 0.135 | 0.118 | 0.086 | 0.093 | 0.101 | 0.109 | 0.093 | 0.110 | 0.112 | 0.116 | 0.118 | 0.124 | 0.135 |
| Liberia | 0.050 | 0.018 | 0.018 | 0.018 | 0.018 | 0.024 | 0.034 | 0.041 | 0.050 | 0.070 | 0.073 | 0.073 | 0.034 | 0.024 | 0.073 |
| Libyan Arab Jamahirya | 0.029 | 0.029 | 0.028 | 0.028 | 0.028 | 0.036 | 0.038 | 0.033 | 0.028 | 0.028 | 0.029 | 0.028 | 0.028 | 0.028 | 0.028 |
| Madagascar | 0.007 | 0.009 | 0.009 | 0.010 | 0.014 | 0.014 | 0.031 | 0.018 | 0.020 | 0.022 | 0.027 | 0.030 | 0.031 | 0.037 | 0.009 |
| Malawi | 0.061 | 0.064 | 0.064 | 0.071 | 0.064 | 0.080 | 0.064 | 0.071 | 0.061 | 0.064 | 0.064 | 0.080 | 0.064 | 0.064 | 0.071 |
| Malaysia | 0.650 | 0.706 | 0.717 | 0.720 | 0.733 | 0.729 | 0.744 | 0.755 | 0.778 | 0.819 | 0.817 | 0.775 | 0.769 | 0.765 | 0.762 |
| Maldives | 0.370 | 0.375 | 0.370 | 0.370 | 0.375 | 0.306 | 0.306 | 0.380 | 0.306 | 0.357 | 0.344 | 0.379 | 0.385 | 0.370 | 0.375 |
| Malta | 0.848 | 0.862 | 0.892 | 0.955 | 0.979 | 1.000 | 1.000 | 1.000 | 1.000 | 1.000 | 1.000 | 1.000 | 1.000 | 1.000 | 1.000 |
| Mauritania | 0.042 | 0.052 | 0.052 | 0.042 | 0.047 | 0.056 | 0.042 | 0.042 | 0.041 | 0.046 | 0.043 | 0.047 | 0.052 | 0.056 | 0.056 |
| Mauritius | 0.541 | 0.586 | 0.570 | 0.598 | 0.625 | 0.640 | 0.652 | 0.672 | 0.679 | 0.691 | 0.661 | 0.649 | 0.609 | 0.625 | 0.618 |
| Mexico | 0.127 | 0.129 | 0.151 | 0.156 | 0.247 | 0.253 | 0.288 | 0.226 | 0.236 | 0.310 | 0.226 | 0.237 | 0.255 | 0.236 | 0.248 |
| Micronesia, Federated States of | 0.124 | 0.125 | 0.129 | 0.127 | 0.128 | 0.141 | 0.149 | 0.155 | 0.156 | 0.157 | 0.159 | 0.128 | 0.120 | 0.114 | 0.111 |
| Moldova | 0.247 | 0.267 | 0.288 | 0.328 | 0.356 | 0.362 | 0.379 | 0.390 | 0.403 | 0.429 | 0.416 | 0.288 | 0.327 | 0.348 | 0.365 |
| Mongolia | 0.107 | 0.129 | 0.139 | 0.216 | 0.232 | 0.238 | 0.271 | 0.302 | 0.330 | 0.376 | 0.409 | 0.427 | 0.529 | 0.451 | 0.488 |
| Montenegro | 0.593 | 0.578 | 0.537 | 0.335 | 0.578 | 0.535 | 0.537 | 0.573 | 0.596 | 0.544 | 0.560 | 0.506 | 0.593 | 0.496 | 0.588 |
| Morocco | 0.205 | 0.175 | 0.170 | 0.180 | 0.222 | 0.248 | 0.292 | 0.303 | 0.313 | 0.323 | 0.333 | 0.312 | 0.334 | 0.366 | 0.375 |
| Mozambique | 0.039 | 0.015 | 0.021 | 0.025 | 0.028 | 0.033 | 0.034 | 0.039 | 0.042 | 0.058 | 0.061 | 0.072 | 0.084 | 0.075 | 0.076 |
| Myanmar | 0.027 | 0.028 | 0.028 | 0.029 | 0.030 | 0.031 | 0.032 | 0.030 | 0.035 | 0.038 | 0.043 | 0.055 | 0.062 | 0.056 | 0.073 |
| Namibia | 0.098 | 0.108 | 0.133 | 0.141 | 0.212 | 0.234 | 0.241 | 0.260 | 0.255 | 0.281 | 0.327 | 0.368 | 0.416 | 0.327 | 0.098 |
| Nepal | 0.253 | 0.218 | 0.142 | 0.218 | 0.253 | 0.218 | 0.162 | 0.253 | 0.120 | 0.125 | 0.142 | 0.162 | 0.181 | 0.218 | 0.253 |
| Netherlands | 0.562 | 0.612 | 0.609 | 0.610 | 0.609 | 0.631 | 0.632 | 0.600 | 0.571 | 0.587 | 0.649 | 0.591 | 0.580 | 0.562 | 0.588 |
| Nicaragua | 0.106 | 0.117 | 0.114 | 0.133 | 0.136 | 0.114 | 0.103 | 0.111 | 0.109 | 0.114 | 0.128 | 0.140 | 0.158 | 0.173 | 0.150 |
| North Macedonia | 0.716 | 0.564 | 0.710 | 0.458 | 0.564 | 0.614 | 0.634 | 0.629 | 0.647 | 0.664 | 0.689 | 0.710 | 0.716 | 0.718 | 0.733 |
| Norway | 0.285 | 0.233 | 0.265 | 0.272 | 0.331 | 0.275 | 0.272 | 0.278 | 0.275 | 0.286 | 0.284 | 0.286 | 0.285 | 0.283 | 0.281 |
| Oman | 0.156 | 0.169 | 0.183 | 0.201 | 0.219 | 0.228 | 0.238 | 0.259 | 0.251 | 0.275 | 0.264 | 0.266 | 0.267 | 0.258 | 0.262 |
| Pakistan | 0.047 | 0.050 | 0.062 | 0.066 | 0.067 | 0.063 | 0.068 | 0.073 | 0.077 | 0.080 | 0.086 | 0.091 | 0.094 | 0.100 | 0.102 |
| Panama | 0.310 | 0.493 | 0.315 | 0.325 | 0.315 | 0.310 | 0.353 | 0.375 | 0.412 | 0.449 | 0.476 | 0.493 | 0.508 | 0.513 | 0.518 |
| Papua New Guinea | 0.119 | 0.119 | 0.095 | 0.093 | 0.116 | 0.094 | 0.122 | 0.100 | 0.122 | 0.121 | 0.110 | 0.101 | 0.108 | 0.122 | 0.126 |
| Paraguay | 0.013 | 0.013 | 0.016 | 0.015 | 0.019 | 0.026 | 0.031 | 0.037 | 0.043 | 0.047 | 0.051 | 0.059 | 0.063 | 0.076 | 0.130 |
| Peru | 0.200 | 0.131 | 0.151 | 0.182 | 0.222 | 0.218 | 0.248 | 0.282 | 0.290 | 0.317 | 0.352 | 0.380 | 0.411 | 0.421 | 0.447 |
| Philippines | 0.086 | 0.092 | 0.094 | 0.092 | 0.098 | 0.101 | 0.117 | 0.131 | 0.118 | 0.126 | 0.129 | 0.135 | 0.141 | 0.144 | 0.156 |
| Poland | 0.731 | 0.769 | 0.724 | 0.769 | 0.731 | 0.769 | 0.735 | 0.724 | 0.700 | 0.696 | 0.712 | 0.738 | 0.756 | 0.781 | 0.823 |
| Portugal | 0.759 | 0.759 | 0.759 | 0.810 | 0.827 | 0.824 | 0.830 | 0.828 | 0.751 | 0.759 | 0.762 | 0.765 | 0.712 | 0.697 | 0.711 |
| Romania | 0.083 | 0.084 | 0.083 | 0.092 | 0.099 | 0.099 | 0.095 | 0.090 | 0.090 | 0.086 | 0.083 | 0.083 | 0.083 | 0.084 | 0.092 |
| Rwanda | 0.002 | 0.002 | 0.002 | 0.006 | 0.047 | 0.055 | 0.068 | 0.075 | 0.086 | 0.079 | 0.069 | 0.058 | 0.066 | 0.063 | 0.104 |
| Samoa | 0.142 | 0.175 | 0.184 | 0.181 | 0.204 | 0.211 | 0.211 | 0.316 | 0.337 | 0.416 | 0.382 | 0.396 | 0.477 | 0.481 | 0.482 |
| San Marino | 0.924 | 0.961 | 0.899 | 0.973 | 0.985 | 1.000 | 0.926 | 0.895 | 0.887 | 0.942 | 0.964 | 0.829 | 0.830 | 0.826 | 0.811 |
| Sao Tome and Principe | | | | | | | | 0.203 | 0.239 | 0.279 | 0.257 | 0.330 | 0.254 | 0.271 | 0.270 |
| Saudi Arabia | 0.366 | 0.366 | 0.259 | 0.366 | 0.366 | 0.254 | 0.259 | 0.257 | 0.258 | 0.302 | 0.304 | 0.315 | 0.336 | 0.356 | 0.366 |
| Serbia | 0.057 | 0.057 | 0.044 | 0.052 | 0.056 | 0.053 | 0.056 | 0.057 | 0.159 | 0.158 | 0.162 | 0.177 | 0.182 | 0.183 | 0.184 |
| Seychelles | 0.356 | 0.426 | 0.438 | 0.430 | 0.436 | 0.463 | 0.462 | 0.481 | 0.561 | 0.537 | 0.572 | 0.555 | 0.572 | 0.608 | 0.621 |
| Solomon Islands | 0.078 | 0.080 | 0.073 | 0.073 | 0.086 | 0.085 | 0.083 | 0.081 | 0.085 | 0.070 | 0.120 | 0.128 | 0.132 | 0.138 | 0.120 |
| South Africa | 0.194 | 0.220 | 0.263 | 0.285 | 0.283 | 0.316 | 0.334 | 0.379 | 0.448 | 0.486 | 0.498 | 0.483 | 0.495 | 0.506 | 0.448 |
| South Sudan | | | | | 0.022 | 0.022 | 0.015 | 0.005 | 0.007 | 0.010 | 0.015 | 0.020 | 0.027 | 0.024 | 0.022 |
| Spain | 0.660 | 0.687 | 0.688 | 0.713 | 0.709 | 0.730 | 0.731 | 0.709 | 0.693 | 0.704 | 0.700 | 0.680 | 0.663 | 0.703 | 0.693 |
| Suriname | 0.479 | 0.460 | 0.479 | 0.407 | 0.479 | 0.479 | 0.479 | 0.446 | 0.446 | 0.446 | 0.479 | 0.407 | 0.460 | 0.446 | 0.479 |
| Sweden | 1.000 | 1.000 | 1.000 | 0.976 | 0.976 | 1.000 | 0.925 | 0.945 | 0.976 | 1.000 | 1.000 | 1.000 | 1.000 | 1.000 | 1.000 |
| Switzerland | 0.743 | 0.740 | 0.724 | 0.715 | 0.697 | 0.726 | 0.721 | 0.735 | 0.731 | 0.788 | 0.796 | 0.729 | 0.635 | 0.621 | 0.589 |
| Syrian Arab Republic | 0.075 | 0.069 | 0.069 | 0.074 | 0.069 | 0.074 | 0.075 | 0.076 | 0.073 | 0.073 | 0.073 | 0.075 | 0.069 | 0.074 | 0.073 |
| Tanzania | 0.044 | 0.060 | 0.038 | 0.044 | 0.048 | 0.044 | 0.052 | 0.058 | 0.060 | 0.060 | 0.108 | 0.071 | 0.071 | 0.048 | 0.044 |
| Thailand | 0.313 | 0.340 | 0.370 | 0.396 | 0.404 | 0.405 | 0.418 | 0.416 | 0.442 | 0.457 | 0.470 | 0.478 | 0.484 | 0.497 | 0.519 |
| Tonga | 0.180 | 0.261 | 0.312 | 0.201 | 0.191 | 0.173 | 0.161 | 0.169 | 0.202 | 0.233 | 0.260 | 0.309 | 0.247 | 0.321 | 0.362 |
| Trinidad and Tobago | | | | | 0.501 | 0.516 | 0.529 | 0.533 | 0.536 | 0.542 | 0.537 | 0.537 | 0.536 | 0.515 | 0.509 |
| Turkey | 0.417 | 0.417 | 0.424 | 0.442 | 0.417 | 0.422 | 0.730 | 0.751 | 0.781 | 0.847 | 0.820 | 0.841 | 0.899 | 0.980 | 1.000 |
| Uganda | 0.028 | 0.028 | 0.034 | 0.038 | 0.043 | 0.048 | 0.052 | 0.049 | 0.055 | 0.052 | 0.056 | 0.067 | 0.073 | 0.100 | 0.130 |
| Ukraine | 0.870 | 0.809 | 0.870 | 0.943 | 0.946 | 0.988 | 1.000 | 1.000 | 1.000 | 0.731 | 0.746 | 0.716 | 0.699 | 0.742 | 0.615 |
| United Arab Emirates | 0.381 | 0.388 | 0.381 | 0.369 | 0.383 | 0.369 | 0.381 | 0.383 | 0.378 | 0.355 | 0.388 | 0.425 | 0.445 | 0.469 | 0.457 |
| United Kingdom | 0.036 | 0.035 | 0.034 | 0.034 | 0.034 | 0.034 | 0.030 | 0.028 | 0.027 | 0.027 | 0.036 | 0.036 | 0.036 | 0.036 | 0.028 |
| Uzbekistan | 0.095 | 0.130 | 0.127 | 0.142 | 0.159 | 0.164 | 0.187 | 0.183 | 0.188 | 0.216 | 0.247 | 0.272 | 0.304 | 0.351 | 0.350 |
| Vanuatu | 0.223 | 0.239 | 0.239 | 0.239 | 0.239 | 0.223 | 0.226 | 0.239 | 0.239 | 0.239 | 0.226 | 0.239 | 0.243 | 0.223 | 0.223 |
| Venezuela, Republica Bolivariana de | 0.111 | 0.054 | 0.076 | 0.110 | 0.113 | 0.118 | 0.111 | 0.118 | 0.125 | 0.163 | 0.190 | 0.259 | 0.113 | 0.259 | 0.125 |
| Vietnam | 0.164 | 0.266 | 0.233 | 0.233 | 0.164 | 0.233 | 0.061 | 0.122 | 0.149 | 0.164 | 0.189 | 0.207 | 0.233 | 0.233 | 0.266 |
| West Bank and Gaza | 0.262 | 0.335 | 0.273 | 0.273 | 0.262 | 0.261 | 0.273 | 0.312 | 0.321 | 0.316 | 0.312 | 0.320 | 0.329 | 0.335 | 0.351 |
| Yemen, Republic of | 0.031 | 0.031 | 0.031 | 0.026 | 0.029 | 0.031 | 0.028 | 0.026 | 0.029 | 0.030 | 0.031 | 0.030 | 0.031 | 0.031 | 0.030 |
| Zambia | 0.077 | 0.077 | 0.076 | 0.082 | 0.077 | 0.076 | 0.036 | 0.041 | 0.052 | 0.071 | 0.079 | 0.082 | 0.083 | 0.086 | 0.077 |
| Zimbabwe | 0.129 | 0.129 | 0.129 | 0.053 | 0.055 | 0.025 | 0.044 | 0.038 | 0.051 | 0.062 | 0.042 | 0.041 | 0.044 | 0.122 | 0.187 |

Source: Own estimates using Financial Access Survey, 2004-2018 (IMF, 2019). A few countries are missing in these tables as some of their information was not available at the time of writing this paper.

Table A.4. *Growth-redistribution decomposition of poverty: the role of financial inclusion (variables in levels)*

| Dependent variable | Headcount | | | Poverty gap | | | Poverty gap squared | | | Watts | | |
|---|---|---|---|---|---|---|---|---|---|---|---|---|
| Financial inclusion type→ | Financial inclusion index | Financial outreach | Usage | Financial inclusion index | Financial outreach | Usage | Financial inclusion index | Financial outreach | Usage | Financial inclusion index | Financial outreach | Usage |
| | 1 | 2 | 3 | 4 | 5 | 6 | 7 | 8 | 9 | 10 | 11 | 12 |
| Gini | 0.807*** | 0.803** | 0.786*** | 0.551*** | 0.553*** | 0.524*** | 0.397*** | 0.398*** | 0.375*** | 1.071*** | 1.072*** | 1.016*** |
| | [0.290] | [0.306] | [0.257] | [0.154] | [0.158] | [0.139] | [0.098] | [0.099] | [0.088] | [0.264] | [0.268] | [0.238] |
| GDP growth rate | 0.181 | 0.17 | 0.204* | 0.096 | 0.094 | 0.101* | 0.057 | 0.058 | 0.059 | 0.154 | 0.154 | 0.158 |
| | [0.116] | [0.118] | [0.116] | [0.059] | [0.059] | [0.059] | [0.036] | [0.036] | [0.036] | [0.098] | [0.098] | [0.099] |
| Financial inclusion | 0.175 | 0.246 | 0.074 | 0.304* | 0.365* | 0.179 | 0.265** | 0.306** | 0.164** | 0.660** | 0.765** | 0.409** |
| | [0.334] | [0.380] | [0.256] | [0.171] | [0.190] | [0.122] | [0.108] | [0.120] | [0.076] | [0.287] | [0.322] | [0.202] |
| Gini x Financial inclusion | -1.24 | -1.65 | -0.594 | -1.058** | -1.256** | -0.621* | -0.821*** | -0.931*** | -0.515** | -2.107** | -2.399*** | -1.313** |
| | [0.928] | [1.015] | [0.754] | [0.477] | [0.518] | [0.353] | [0.303] | [0.331] | [0.217] | [0.805] | [0.882] | [0.575] |
| Constant | -0.03 | -0.025 | -0.038 | -0.111* | -0.112* | -0.104* | -0.102** | -0.103** | -0.094** | -0.263** | -0.266** | -0.245** |
| | [0.120] | [0.125] | [0.106] | [0.063] | [0.064] | [0.057] | [0.040] | [0.040] | [0.036] | [0.107] | [0.110] | [0.097] |
| Observations | 933 | 933 | 933 | 933 | 933 | 933 | 933 | 933 | 933 | 933 | 933 | 933 |
| Adjusted $R^2$ | 0.167 | 0.18 | 0.145 | 0.151 | 0.155 | 0.141 | 0.134 | 0.136 | 0.128 | 0.143 | 0.145 | 0.136 |
| Number of country | 78 | 78 | 78 | 78 | 78 | 78 | 78 | 78 | 78 | 78 | 78 | 78 |
| Country fixed effects | Yes | Yes | Yes | Yes | Yes | Yes | Yes | Yes | Yes | Yes | Yes | Yes |
| Robust standard error cluster | Country | Country | Country | Country | Country | Country | Country | Country | Country | Country | Country | Country |

The dependent variables are *headcount*, *poverty gap*, *poverty gap squared*, and *Watts* index. We use *Gini* as the proxy for income inequality. We use real GDP growth rate. *Financial Inclusion index* is a composite index, constructed based on two dimensions, namely *financial outreach* and *usage* dimensions. ***, **, and * indicate statistical significance at the 1%, 5%, and 10% levels respectively.

Source: World Bank's PovcalNet database and IMF. Coverage: 2004-2018.